\documentclass[
superscriptaddress,
longbibliography,
showpacs,
amsmath,amssymb,
aps,
pra,
twocolumn,
]{revtex4-1}

\usepackage{bibunits}
\usepackage{graphicx}
\usepackage{dcolumn}
\usepackage{bm}
\usepackage[breaklinks,colorlinks = true,linkcolor = red,urlcolor=blue,citecolor=red]{hyperref}
\usepackage{multirow}
\usepackage{array}
\usepackage{booktabs}
\usepackage{ctable}
\usepackage{upgreek}
\usepackage{epsfig}
\usepackage{mathrsfs}
\usepackage{amssymb}
\usepackage{amsbsy}
\usepackage{color}
\usepackage{cancel}
\usepackage{pifont}
\usepackage{marginnote}
\usepackage{float}
\usepackage{verbatim}
\usepackage{subfigure}
\usepackage{bbm, dsfont}
\usepackage{lineno}
\usepackage{amsmath}

\newcommand{\bcen}{\begin{center}}
\newcommand{\ecen}{\end{center}}
\newcommand{\btab}{\begin{tabular}}
\newcommand{\etab}{\end{tabular}}
\newcommand{\bdes}{\begin{description}}
\newcommand{\edes}{\end{description}}

\newcommand{\ul}{\underline}
\newcommand{\beq}{\begin{equation}}
\newcommand{\eeq}{\end{equation}}
\newcommand{\bea}{\begin{eqnarray}}
\newcommand{\eea}{\end{eqnarray}}

\newcommand{\bary}{\begin{array}}
	\newcommand{\eary}{\end{array}}
\newcommand{\benum}{\begin{enumerate}}
	\newcommand{\eenum}{\end{enumerate}}
\newcommand{\bitem}{\begin{itemize}}
	\newcommand{\eitem}{\end{itemize}}

\newcommand{\bsig}{\mbox{\boldmath $ \sigma $}}

\newcommand{\bd} { \mbox{\boldmath $d$}}

\newcommand{\eqn}[1] {eqn.~(\ref{#1})}

\newcommand{\sect}[1] {Section~\ref{#1}}

\newcommand{\Fig}[1]{Fig.~\ref{#1}}

\newcommand{\ci}{\mathbbm{i}}
\makeatletter

\newcommand{\Rmnum}[1]{\expandafter\@slowromancap\romannumeral #1@}
\makeatother

\newcommand{\signum}[0]{\mathop{\mathrm{sign}}}

\newcommand{\titlename}{Chiral metals and entrapped insulators in a one-dimensional topological non-Hermitian system}

\begin{document}

\title{\titlename}

\author{Ayan Banerjee}
\email{ayanbanerjee@iisc.ac.in}
\affiliation{Solid State and Structural Chemistry Unit, Indian Institute of Science, Bangalore 560012, India}
\author{Suraj S. Hegde}
\email{suraj.hegde@tu-dresden.de}
\affiliation{Institut für Theoretische physik, Technische Universität Dresden, 01069 Dresden, Germany}
\affiliation{Max-Planck Institute for the Physics of  Complex Systems, Nöthnitzer straße 38, Dresden 01187, Germany}
\author{Adhip Agarwala}
\email{adhip@pks.mpg.de}
\affiliation{Max-Planck Institute for the Physics of  Complex Systems, Nöthnitzer straße 38, Dresden 01187, Germany}
\affiliation{International Centre for Theoretical Sciences, Tata Institute of Fundamental Research, Bengaluru 560089, India}
\author{Awadhesh Narayan}
\email{awadhesh@iisc.ac.in}
\affiliation{Solid State and Structural Chemistry Unit, Indian Institute of Science, Bangalore 560012, India}

\date{\today}

\begin{abstract}

In this work we study many-body `steady states' that arise in the non-Hermitian generalisation of the non-interacting Su-Schrieffer-Heeger model at a finite density of fermions. We find that the hitherto known phase diagrams for this system, derived from the single-particle gap closings, in fact correspond to distinct non-equilibrium phases, which either carry finite currents or are dynamical insulators where particles are entrapped. Each of these have distinct quasi-particle excitations and steady state correlations and entanglement properties. Looking at finite-sized systems, we further modulate the boundary to uncover the topological features in such steady states -- in particular the emergence of leaky boundary modes. Using a variety of analytical and numerical methods we develop a theoretical understanding of the various phases and their transitions, and uncover the rich interplay of non-equilibrium many-body physics, quantum entanglement and topology in a simple looking, yet a rich model system.

\end{abstract}

\maketitle

\section{Introduction}

Quantum matter and its novel manifestations are often characterized by an intricate interplay of quantum fluctuations, ideas of band topology and entanglement \cite{Wen_RMP_2017,Ludwig_PS_2015, Chiu_RMP_2016,Hasan_RMP_2010, Qi_RMP_2011}. A rather new addition to this group has been the notion of non-Hermiticity, which has its own illustrious history of being visited time and again, in context of dynamical systems, open systems or parity-time ($PT$) symmetric systems, which have been realized in various physical platforms \cite{alvarez2018topological,torres2019perspective,ashida2020non,bergholtz2021exceptional, Carlstrom20,Huber20,Lee14,Lee14a,Lieu20,Wei17,vanCaspel19,Dangel18,Panda20,Lieu_PRA-2019,ghatak2019new,guo2021entanglement,Wang_2021}. While one strategy has been to investigate the role of non-Hermiticity that arises in the effective Hamiltonian of some degrees of freedom (due to coupling to a bath or leads etc.), another has been to sideline the above question and explicitly model the non-Hermitian Hamiltonians and investigate their properties.

A plethora of work, in recent years, has been dedicated to this latter theme and, in particular, the role of non-Hermiticity in lattice models has been widely investigated~\cite{alvarez2018topological,ghatak2019new,torres2019perspective,ashida2020non,bergholtz2021exceptional}. Most notable among this class of works has been the investigations into the role of topology vis-a-vis non-Hermiticity.  While building a catalog of such results, here, would be an impossible task; we point out the underlying themes which have been in discussion in the literature (see References~\cite{alvarez2018topological,ghatak2019new,torres2019perspective,ashida2020non,bergholtz2021exceptional} for recent reviews).  Introduction of non-Hermiticity, in general, in the Hamiltonian matrix results in complex eigenvalues, which leads to the idea of a band diagram being generalized to a complex energy plane, where the notion of a band gap is modified to prohibition of touching a base energy in the complex spectrum \cite{Gong18}. Thus a new notion of `a topological phase' was introduced in these systems, which involves winding of the single particle energies over the entire complex spectra \cite{Gong18, Kawabata19,lee2019anatomy}. It is important to note, however, that such a winding is inherently distinct from the case where a half-filled Hermitian system is called ``topological", where it is often a response of the many-particle state in equilibrium that can be written in terms of such topological quantities which quantify the behaviour of single particle wavefunctions over the Brillouin zone \cite{Resta_RMP_1994, Watanabe_PRX_2018, Thouless_PRL_1982}. Absence of this latter connection is a result of a rather incomplete understanding between the many-body state of a non-Hermitian non-interacting system and its single particle complex eigenspectrum, as we discuss below.

While the object of interest in Hermitian lattice tight-binding models is the ground state and its low energy excitations, in a non-Hermitian system, the appearance of complex eigen-spectra requires us to restructure the framework for characterising the phases of matter. For instance, given a finite density of particles (say, half-filling of bands) in a non-interacting equilibrium system, a gapped single particle spectrum implies that the many-body state is characterized by correlations, which are often topological invariants of the lowest filled band. Such a direct mapping between the single-particle states and the many-body state is absent for non-Hermitian systems, given the system may not reach an equilibrium state but rather a non-equilibrium steady state at long times. Therefore, the method of conventional energy minimization to determine a many-body state is not directly applicable to a non-Hermitian system. In fact, not unlike the classical non-reciprocal transitions, it is possible that the description of the many-body state is eminent only within a set of dynamical equations rather than an underlying free energy \cite{Fruchart2021}.

In this work we study such non-equilibrium steady states that arise in a concrete setting of a microscopic non-Hermitian lattice model at a finite density of fermions. The prescription to form such steady states following References \cite{Lieu_PRA-2019,LieuPRB19, Panda20, Carlstrom20} is the following. Given a non-interacting Hamiltonian which has a set of single particle eigenvalues given by $E_i=E^{R}_i+ \ci E^{I}_i$ ($i=\{1,\ldots, L\}$), a many-body state constructed out of $N_p$ particles is a direct product of such single particle states, with a corresponding many body energy. Note that we use $\ci$ for the square root of unity, while reserving $i$ for the site index. There are $^{L}C_{N_p}$ number of such many-body states defined by $|\Psi_j\rangle$ with energy $\mathcal{E}_j = \mathcal{E}^R+ \ci \mathcal{E}^I$. Each of these states evolves in time with a generalized time-evolution operator. A general many-body wave-function in this Hilbert space will, therefore, evolve as 

\beq
|\Psi(t)\rangle = \sum_j \psi_j |\Psi_j(t)\rangle  = \sum_j \psi_j e^{-\ci(\mathcal{E}_j) t} |\Psi_j (t=0)\rangle.
\eeq

So, any generic many-body state at late times ($t \rightarrow \infty$) will eventually evolve into the many-body eigenstate with the largest value of $\mathcal{E}^I$. {\it We define this direct product state as the steady state of the system}.

\begin{figure}
    \centering
    \includegraphics[width=1.0\columnwidth]{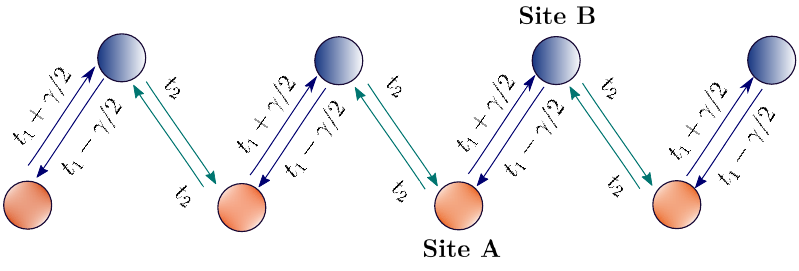}
    \caption{\textbf{Schematic of the non-Hermitian SSH model.} The non-Hermitian SSH model with non-reciprocal hoppings. Intra-unit cell hopping along the right and left are $t_1\pm \gamma/2$, giving rise to non-Hermiticity. Inter-unit cell hopping is fixed at $t_2$.}
    \label{fig:Model}
\end{figure}

The model we focus on in this work is the generalization of the paradigmatic Su-Schrieffer-Heeger (SSH) Hamiltonian but with non-reciprocal hoppings to model the effect of non-Hermiticity (see \Fig{fig:Model} for a schematic). In particular, here, we would discuss the physics of this system in light of the {\it many-body steady states} it realizes -- the correlations, low energy excitations, the entanglement signatures and the effect of tuning the boundary conditions. Given that this is a non-interacting system, not unexpectedly, all the properties can be determined from the single-particle eigenspectrum -- however, as we will show below, this steady states interpretation and the intervening phase transitions provide insights into the nature of distinct non-equilibrium phases and their transitions in non-Hermitian quantum systems. 

The choice of nH-SSH (non-Hermitian SSH) model to illustrate this physics is most natural given the immense body of work that has been dedicated towards this particular model \cite{Lee_PRB_2020,Herviou_PRA_2019,Lieu_PRB_2018,Jin_PRB_2019,yao2018non,kunst2018biorthogonal,liu2019topological,song2019non,LieuPRB19,KentaPRB19,Schomerus13,MartinezPRB18,YinPRA18,YaoPRL18,StJeanNature17,Albrecht_Nature_2016,KunstPRB2019,Dangel18,borgnia2020non,masuda2021relationship, Zirnstein_PRL_2021,vyas2021topological, he2020non, Silberstein_PRB_2020}. In particular, we follow the notations introduced in Reference \cite{Herviou_PRA_2019}, where a phase diagram for the nH-SSH model is presented in terms of the single-particle gap closings both for periodic and open boundary conditions, showing that the bulk-edge correspondence does not translate, in a straight forward manner, to non-Hermitian systems\cite{helbig2020generalized,hofmann2020reciprocal,weidemann2020topological}. Interestingly, in Reference \cite{YaoPRL18}, an interpretation for appearance of boundary modes and related non-Hermitian skin effect was understood within the framework of a generalized Brillouin zone, while in Reference \cite{kunst2018biorthogonal}, such a correspondence was seen with bulk bi-orthogonal polarization.
In this work, we revisit this problem in terms of the many-body steady states the system realizes. We find four distinct non-equillbirium phases with characteristic currents, low-energy excitations, correlations and entanglement signatures. We provide a comprehensive understanding of these non-equilibrium phases and the nature of the transitions between them. We further discuss the effects of weakening the boundary term, such that topological boundary modes appear in these steady states. We further point out various connections between this many-body perspective and the single-particle dispersion framework discussed previously in the literature \cite{Herviou_PRA_2019}.

\begin{figure}
    \centering
    \includegraphics[width=1.0\columnwidth]{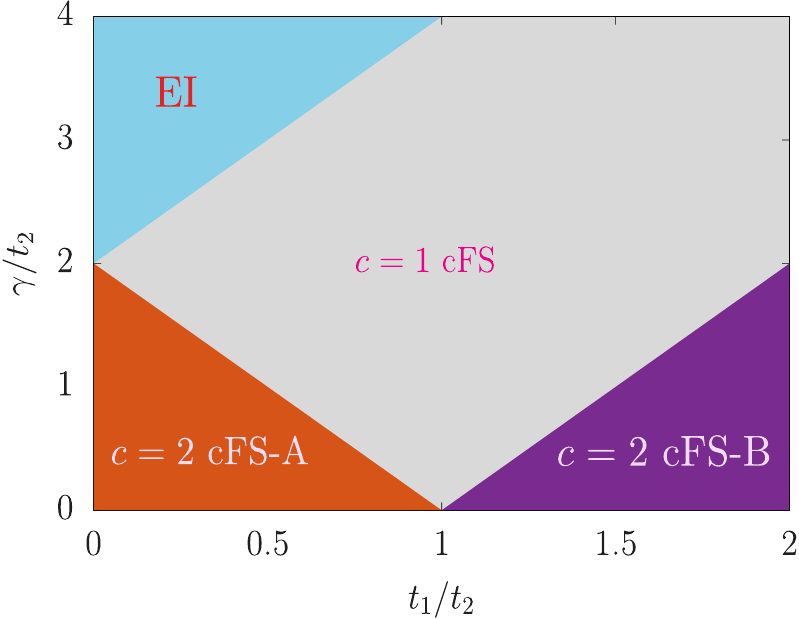}
    \caption{\textbf{Phase diagram of the non-Hermitian SSH model.} The non-equilibrium phases that arise in the non-Hermitian SSH model at half-filling [see \eqn{eqn:Ham}], shows four distinct phases. These include two central charge, $c=2$, chiral-Fermi sea (cFS) phases (labelled A and B), one $c=1$ cFS phase, and a non-Hermitian chiral insulating state with $c=0$, termed entrapped insulator (EI).}
    \label{phasediag}
\end{figure}


The central finding of this work is presented in \Fig{phasediag}, where we show that by varying the parameters characterizing the nH-SSH model -- (a) $t_1/t_2$: the dimerization scale and (b) $\gamma/t_2$: the non-reciprocal intra-unit-cell hopping (see \Fig{fig:Model}), one can access four distinct non-equillbrium phases (i) a chiral metal with central charge $2$ (A) [$c=2$ cFS-A], (ii) a chiral metal with central charge $1$ [$c=1$ cFS], (iii)  another chiral metal with central charge $2$ [c$=2$ cFS-B], and (iv) an entrapped insulator EI [$c=0$].

In the rest of the paper we discuss the physics underlying these phases. In \sect{sec:Model} we introduce the Hamiltonian and discuss its symmetries. In \sect{sec:methodology} we present the methodology and introduce various diagnostic tools to characterize these phases. In \sect{sec:phases} we explore the emergence of these four distinct phases and their properties and connections to spectral topology, as well as their low energy theories. In \sect{sec:boundary} we examine the effect of tuning the boundary conditions from periodic to open and the realization of leaky boundary modes. Finally, in \sect{sec:summary} we summarize our results and present an outlook.

\section{Non-Hermitian SSH model}
\label{sec:Model}

\subsection{Model}

\begin{figure}
\includegraphics[width=1.0\columnwidth]{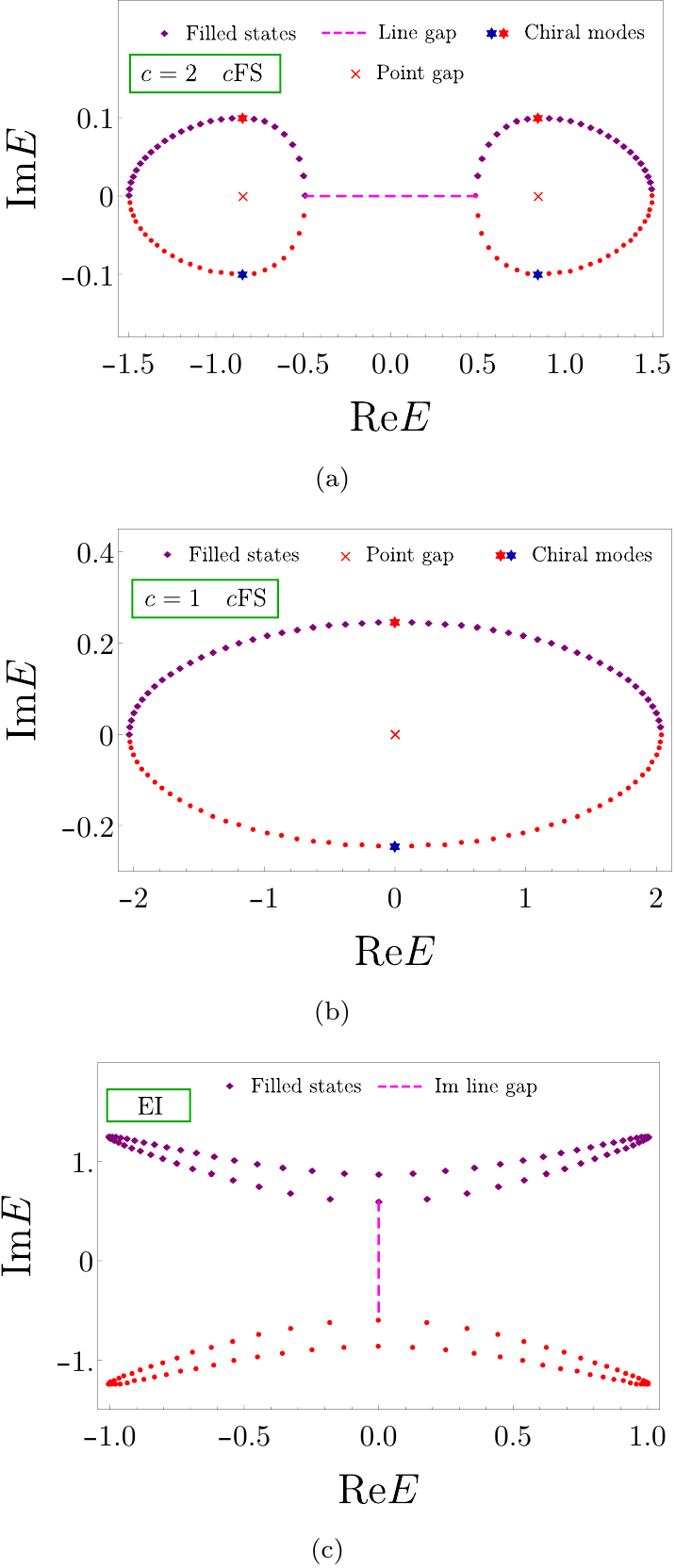}
\caption{\textbf{Nature of complex eigenspectrum.} The real and imaginary parts of the single particle energy eigenvalues are shown in (a) $c=2$ cFS-A ($t_1=0.5$ and $\gamma=0.2$ ) with two spectral loops separated by a real line gap, (b) $c=1$ cFS, ($t_1=1.05$ and $\gamma=0.5$) which has a single loop, and in (c) entrapped insulator phase ($t_1=0.1$ and $\gamma=2.5$) which has two loops but separated by an imaginary line gap. The purple diamonds denote the filled states at half-filling while red (blue) stars show growing (decaying) chiral modes. The dotted lines (crosses) indicate the line gap (point gaps). Here, $t_1$ and $\gamma$ are scaled in terms of $t_2$. We stick to this convention unless otherwise specified. } 
\label{fig:chiral-modes}
\end{figure}


The Hamiltonian of interest for the nH-SSH model is given by~\cite{Herviou_PRA_2019} (also called ``chiral nH-SSH" model, see for e.g.~\cite{Lieu_PRB_2018}), 

\begin{align}
H =  - & \sum_{i} [t_1 (c^\dagger_{i,A} c_{i,B} + h.c.) +
t_2 (c^\dagger_{i+1,A} c_{i,B} + h.c.)]  \notag \\ +& \sum_{i}
\frac{\gamma}{2} (c^\dagger_{i,B}c_{i,A} - c^\dagger_{i,A}c_{i,B}),
\label{eqn:Ham}
\end{align}

where $c^{\dagger}_{i,\alpha} (c_{i,\alpha})$ is the fermionic creation (annihilation) operator at site $i$ for sublattice $\alpha=A,B$ (see \Fig{fig:Model}). The intra- and inter-unit cell hopping amplitudes are given by $t_1$ and $t_2$, respectively, and $\gamma$ introduces a non-reciprocity only in the intra-unit cell hopping, thereby introducing non-Hermiticity in the system. We consider $t_1,t_2,\gamma$ to be real valued. Moving to $k$-space by Fourier transforming the operators [using $c^\dagger_{k,A} =  \frac{1}{\sqrt{L}} \sum_i \exp(-\ci k i) c^\dagger_{i,A}$], we get

\begin{equation} 
  \begin{split}
H(k)  =\bd(k)\cdot \bsig=
\begin{bmatrix}
0 & t_2 e^{-\ci k}+t_1-\gamma/2\\
t_2 e^{\ci k}+t_1+\gamma/2 & 0
\end{bmatrix},
\end{split}
\label{specexpan}
\end{equation}      

where $\bd(k) \equiv (d_1, d_2, d_3)=(t_1+t_2 \cos{k},t_2\sin{k}-\ci\gamma/2,0)$ and $\bsig$ is the triad of Pauli matrices. The eigenvalues of $H(k)$ are given by $E_k=\pm d(k)$, where

\begin{equation}
d(k)=\pm \sqrt{t_1^2+t_2^2 -\gamma^2/4 + 2 t_1 t_2\cos{k}-\ci t_2 \gamma\sin{k}}.
\end{equation}

The right eigenvectors (corresponding to $\pm d(k)$ energy) are of the form

\begin{equation}
|\psi^{\pm}\rangle=\dfrac{1}{\sqrt{1+|a|^2}}\begin{pmatrix}
  a\\
   \pm 1 \\
  \end{pmatrix},
  \label{wf}
\end{equation}

where

\beq 
a=-\sqrt{\frac{t_1+t_2e^{-\ci k}-\gamma/2}{t_1+t_2e^{\ci k}+\gamma/2} }.
\eeq

The Hamiltonian has a sub-lattice symmetry ${\cal S}: c_{iA} \rightarrow -c^\dagger_{iA}, c_{iB} \rightarrow c^\dagger_{iB}, ({\cal S}  H^\dagger {\cal S}^{-1} = H )$, $\sigma_zH(k)\sigma_z^{-1}=-H(k)$ \cite{Kawabata19}.

\subsection{Single particle eigenvalues and spectral topology}

The nH-SSH model has been extensively studied. Here we briefly summarize the main results (see for e.g.,\cite{Herviou_PRA_2019,Lieu_PRA-2019}). The single particle spectrum shows gap closings (for absolute values of energies) for lines $\gamma= 2(t_1 \pm 1)$ and $\gamma=2(1-t_1)$ dividing the phase diagram into four regions (see \Fig{phasediag}): (i) $\gamma < 2(1-t_1)$ -- the complex spectrum when plotted in the complex plane $\{Re[E], Im[E]\}$ comprises of two spectral lobes separated in real energy by a line gap \cite{Kawabata19,YinPRA18,ghatak2019new} [see \Fig{fig:chiral-modes}(a)], (ii) $t_1 > 1+ \gamma/2$ -- spectral topology is again similar to (i) [see \Fig{fig:chiral-modes}(a)], (iii) $\gamma > 2(|1-t_1|), \gamma< 2(1+t_1)$ -- the spectral topology is that of a single loop in the complex plane [see \Fig{fig:chiral-modes}(b)], and (iv)  $\gamma>2(1+t_1)$ -- the complex spectrum has a two lobe structure with the lobes vertically displaced along the imaginary axis [see 
\Fig{fig:chiral-modes}(c)]. 

The spectral topology can be characterized by the winding number, $w$, \cite{Kawabata19}

\begin{equation}
    w=\int_{-\pi}^{\pi} \frac{dk}{2\pi \ci} \partial_k (\ln(\text{det}[H(k)-E_B])),
\end{equation}

where $E_B$ is some base energy. Immediately, the four regions [(i)-(iv)] are characterized by winding numbers $(1,1,1/2,0)$ \cite{Kawabata19,YinPRA18,ghatak2019new} respectively. Interestingly, these winding numbers have a one-to-one correspondence with non-decaying chiral modes, as we briefly discuss next.

\begin{figure}
\includegraphics[width=0.75\columnwidth]{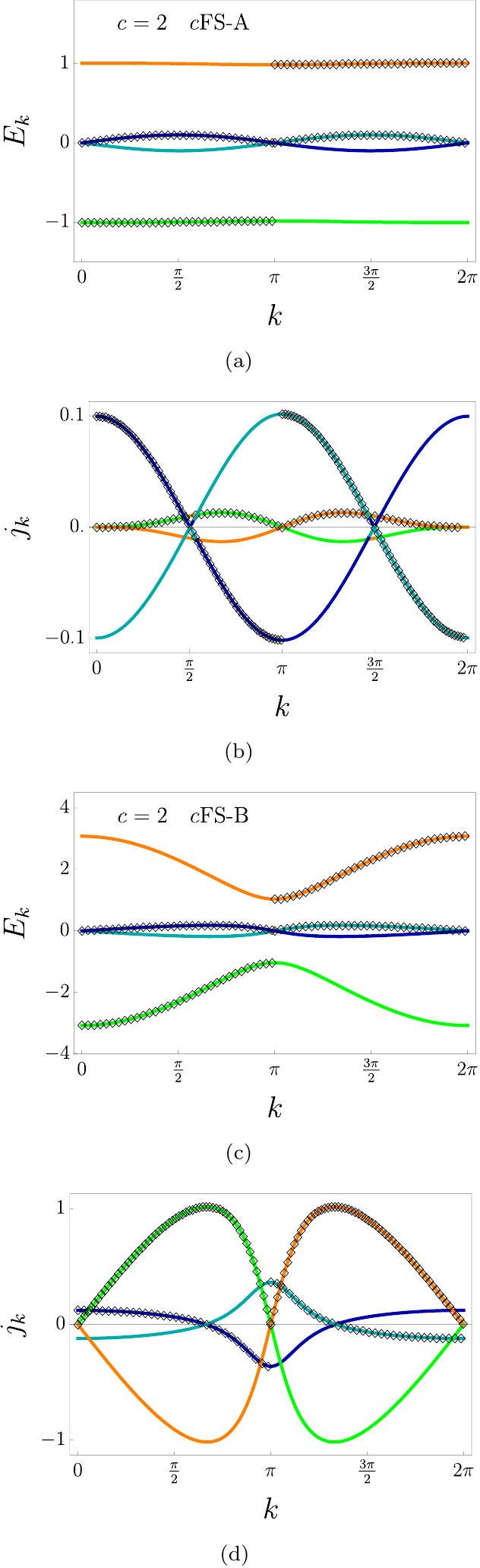}
\caption{\textbf{Single particle dispersion and currents of $c=2$ chiral metals.} The real (orange-green lines) and imaginary part (cyan-blue lines) of the eigenspectrum and single particle current [see \eqn{singlePcurrent}] as a function of $k$ for various parameters are shown. In (a-b) $t_1=0.01,\gamma=0.2$ [$c=2$ cFS-A] (c-d) $t_1=2.1,\gamma=0.75$ [$c=2$ cFS-B]. For half-filling, the occupied $k$ points are further highlighted by black diamonds.}
   \label{fig:dispersion1}
\end{figure}

Each of the single particle states are associated with a current determined by the group velocity ($\sim \frac{d Re[E_k]}{dk}$) and a lifetime given by the inverse of imaginary part of the energy eigenvalue $Im[E_k]$. This immediately shows the existence of chiral modes which decay/grow in the single particle spectrum (see \Fig{fig:chiral-modes}). In an equilibrium system, such modes are stationary states at equal energy which guarantees that any equilibrium state cannot have a finite current (this follows from the celebrated Nielsen-Ninomiya theorem \cite{NIELSEN1981219}). However in a non-equilibrium setting, such as here, these chiral modes have imaginary part of energy eigenvalue of opposite sign, which means that at long times, only one can be populated leading to steady states which can carry a finite current \cite{bessho2020}.  In fact, the existence of such chiral modes and the finite winding numbers are inter-related, symmetry protected and guaranteed via the properties of the spectral topology \cite{Kawabata19,LeePRL2019,yoshida2021correlation}.
Having discussed the properties of the nH-SSH model using its single-particle spectrum and its properties, we next delve into the discussion of the many-body states obtained here at a {\it finite filling}.


\begin{figure}
\includegraphics[width=0.75\columnwidth]{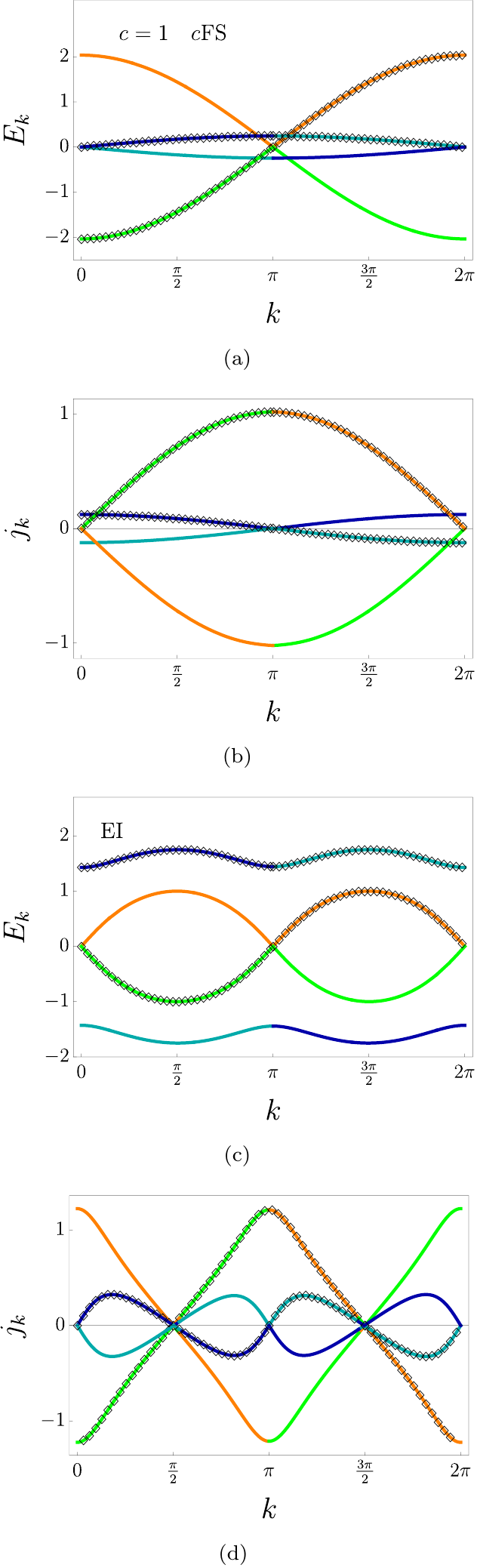}
\caption{\textbf{Single particle dispersions and currents of $c=1$ cFS and $c=0$ EI.}
The real (orange-green lines) and imaginary part (cyan-blue lines) of the eigenspectrum and single particle current [see \eqn{singlePcurrent}] as a function of $k$ for various parameters are shown. In (a-b) $t_1=1.05,\gamma=0.5$ [$c=1$ cFS], and (c-d) $t_1=0.01,\gamma=3.5$ [EI]. For half-filling, the occupied $k$ points are further highlighted by black diamonds.}
\label{fig:dispersion2}
\end{figure}


\section{Methodology}
\label{sec:methodology}

In this section we briefly discuss our method for construction of steady states at finite filling, and define quantities of interest such as current and correlation functions in these steady states.
 
A quadratic non-Hermitian Hamiltonian of the form ${\cal H} = \Psi^\dagger H \Psi$ is diagonalized using $H = U_R \Lambda U^{-1}_R$ where 

\bea
\tilde{c}^\dagger_i = \sum_{j=1}^L (U_R)_{ji} c^\dagger_j \qquad
\tilde{c}_i = \sum_{j=1}^L (U^{-1}_R)_{ij} c_j
\label{eqn:transformation}
\eea

and $c^\dagger_j (c_j)$ are the fermionic creation (annihilation) operators defined at site $j$ such that ${\cal H} = \sum_i E_i \tilde{c}^\dagger_i \tilde{c}_i$ 
where $E_i =E^{R}_i+ \ci E^{I}_i$, and $E^{R}_i$ and $E^{I}_i$ are the real and imaginary parts respectively of every eigenvalue $E_i$. These operators satisfy the conventional fermionic algebra $\{ \tilde{c}^\dagger_i, \tilde{c}_j \} =  \{ c^\dagger_k, c_l \} = \delta_{ij}$ and $\{ \tilde{c}^\dagger_i, \tilde{c}^\dagger_j \} =    \{ c^\dagger_k, c^\dagger_l \} = 0$, such that any particle number state with $N_p$ particles 

\beq
|\Psi_{\alpha} \rangle = \tilde{c}^\dagger_{i_1} \tilde{c}^\dagger_{i_2} \ldots \tilde{c}^\dagger_{i_{N_p-1}} \tilde{c}^\dagger_{N_p} | \Omega \rangle,
\eeq

has a well-defined many-body energy ${\cal H}|\Psi_{\alpha} \rangle =  \Big( \sum_{j=1}^{N_p} E_{i_j}  \equiv \mathcal{E}_{\alpha} \Big) |\Psi_{\alpha} \rangle$, that evolves in time as

\beq
|\Psi_\alpha (t)\rangle  = e^{-\ci(\mathcal{E}_{\alpha}) t} |\Psi_{\alpha} (t=0)\rangle.
\eeq

Given the non-unitary nature of evolution, such a many-body state needs to be renormalized during the time evolution \cite{Faisal_JPB_1981}.
The state which grows the fastest is given by maximum value of $\mathcal{E}^I_\alpha $. Therefore, at long times, the steady state of the system is defined as the one which has the largest value of $\mathcal{E}^I_\alpha$. Any other many-body state ($|\Psi_{\alpha'}\rangle$) with energy $\mathcal{E}^R_{\alpha'}+ \ci \mathcal{E}^I_{\alpha'}$ has a real gap $\Delta^R_{\alpha \alpha'} =\mathcal{E}^R_{\alpha'}-\mathcal{E}^R_{\alpha}$ and similarly an imaginary gap as $\Delta^I_{\alpha \alpha'} =\mathcal{E}^I_{\alpha'}-\mathcal{E}^I_{\alpha}$. The latter also defines the lifetime of the $\alpha'$ state $\equiv \frac{1}{\Delta^I_{\alpha \alpha'}}$  for the excitation to decay into the steady state. Applying this for the nH-SSH model, we calculate the following quantities:

\paragraph{Current response:}{

The current in any system is the response of the Hamiltonian to an external gauge field. Threading a small flux $\phi$ [$t_2 \rightarrow t_2 e^{\ci\frac{\phi}{L}}$] through a one-dimensional ring [see \eqn{eqn:Ham}] leads to the current operator as 

\beq
J = \lim_{\phi \rightarrow 0} -L\frac{\partial H(\phi)}{\partial \phi} =- \ci\sum_{i} \left(  t_2 c^\dagger_{i+1,A}c_{i,B} - t_2  c^\dagger_{i,B} c_{i+1,A} \right).
\label{Current}
\eeq

When Fourier transformed (using $c^\dagger_{k,A} =  \frac{1}{\sqrt{L}} \sum_i \exp(-\ci k i) c^\dagger_{i,A}$), the operator becomes 

\beq
J = -\sum_{k} \ci e^{\ci k} c^\dagger_{k,A}c_{k,B} + \ci e^{ -\ci k} c^\dagger_{k,B} c_{k,A}.
\eeq

When evaluated on a single particle state $|\psi^{\pm}\rangle$ [see \eqn{wf}], one obtains

\beq
j_{k\pm} = \pm \frac{\ci}{2 L} (-e^{-\ci k}/a+e^{\ci k} a ).
\label{singlePcurrent}
\eeq

For a filled state, depending on the occupied states one can, therefore, straightforwardly calculate the total current which is an integral over the corresponding region of the Brillouin zone.}

The total current, $J$, can be equivalently expressed as follows
\begin{equation}
    J= \int_0^{2\pi} n(E_k) \langle j_k \rangle dk= \oint n(E_k) dE_k,
    \label{Eq:TopCurrent}
\end{equation}
where the integral is over the spectral loop in the complex energy plane and $n(E_k)$ is an effective distribution function which defines which states are occupied  \cite{Zhang_2020}. Therefore the way the steady states are formed here is to consider a distribution function of the kind 
\beq
n(E_k) = \frac{1}{1+\exp(-\beta(\text{Im}[E_k]-\mu))},
\eeq

where $\beta=1/T$ and $\mu$ is the chemical potential which is set to zero for the half-filled case.

\paragraph{Correlation functions:}{
Given the occupancy of the single-particle states, the correlation functions are given by 

\bea 
 C^{B,A}(r) &=& \frac{1}{L} \sum_{k} \frac{1}{2a} e^{-\ci kr} (  n_{k +}-n_{k -}), \\
C^{A,B}(r) &=& \frac{1}{L} \sum_{k} \frac{a}{2} e^{-\ci kr} (  n_{k +}-n_{k -}), \\
C^{A,A}(r) &=& \frac{1}{L} \sum_{k} \frac{1}{2} e^{-\ci kr} (  n_{k +}+n_{k -}),
\eea
where $n_{k \pm}$ are the occupations of the $\pm$ bands.

\paragraph{Entanglement:}{Having calculated the correlators, it is straightforward to evaluate the entanglement entropy, $S_l$, of a sub-system of length ($l$) \cite{Peschel_JPA_2003}. The eigenvalues of the subsystem  limited correlator matrix ($\equiv e_i$) are related to the subsystem entanglement entropy via 

\beq
S_l = -\sum_i \Big( e_i \log (e_i) + (1-e_i) \log (1-e_i) \Big).
\eeq

\begin{figure}
\includegraphics[width=1.0\columnwidth]{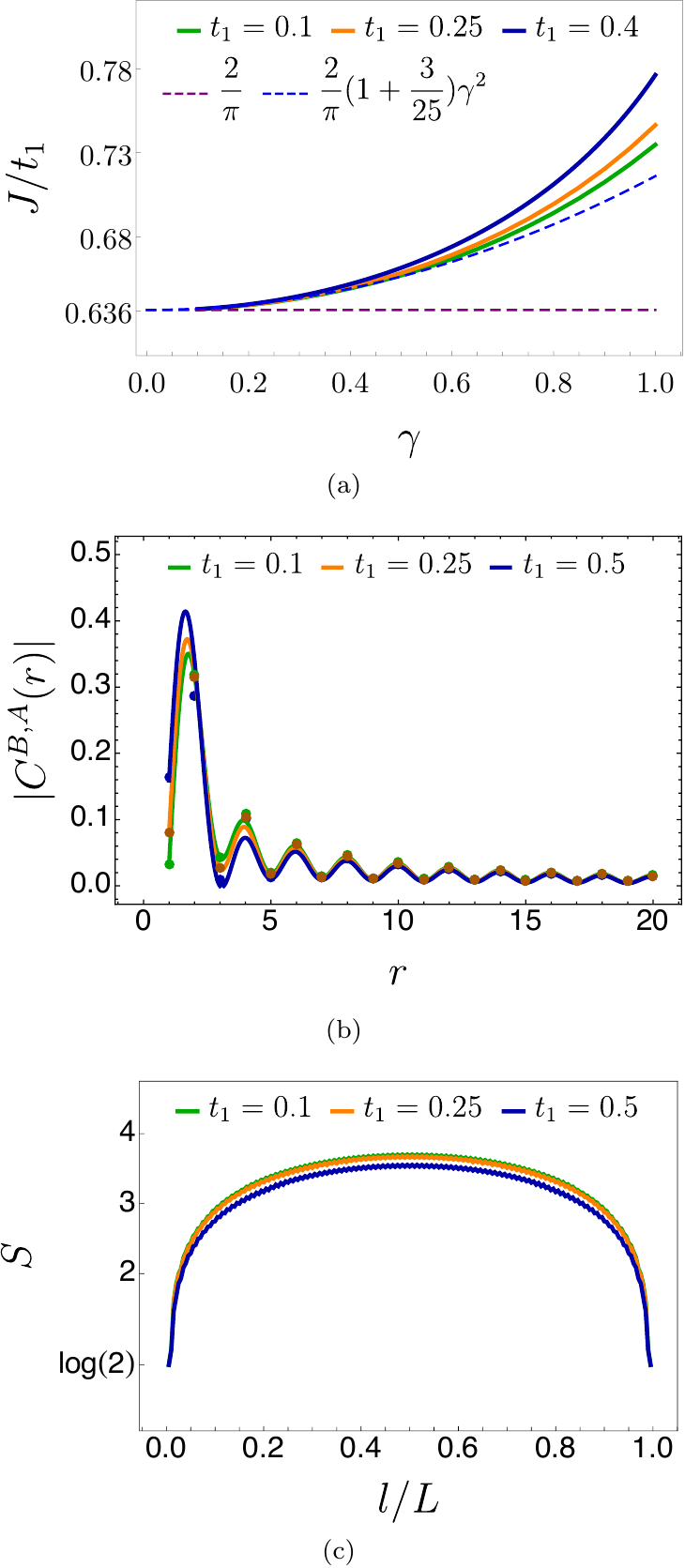}
   \caption{\textbf{Nature of $c=2$ cFS-A.} (a) The behavior of chiral current, for three different values of $t_1$, as a function of the non-Hermitian co-efficient $\gamma$ [see \eqn{c=2-current}]. (b) The correlator, $C_{B,A}(r)$, as a function of $r$ at $\gamma=0.5$ for different values of $t_1$. The numerical values from real space calculation are the points ($L=300$). The continuous lines are the expressions shown in \eqn{corr-cfs-2} for appropriate parameters. (c) Entanglement entropy for three different values of $t_1$ as a function of $l/L$ for $L = 300$.}
   \label{fig:cfs-A}
\end{figure}

Having discussed the various quantities of interest, we next investigate the various non-equilibrium phases realized in the nH-SSH model.

\section{Non-equilibrium phases and their transitions}
\label{sec:phases}


The non-equilibrium phase diagram as a function of $t_1/t_2$ and $\gamma/t_2$, that we obtain, is shown in \Fig{phasediag}. We discover four distinct non-equilibrium phases which we describe below. Before introducing the non-Hermiticity parameter $\gamma$, let us briefly recall the phases which arise as a function of $t_1$ at $\gamma=0$.

\paragraph{\ul{ $\gamma=0$ limit:}} At $\gamma=0$ for both $t_1<t_2$ and $t_1>t_2$ one has gapped phases, which have a bond dimer like order. The fermion-fermion correlator decays exponentially with a functional form $C_{i,j}\sim e^{-r/\xi}$, where $\xi$ is the correlation length $(\xi^{-1}=\dfrac{1}{2}|\ln{\frac{t_1}{t_2}}|)$ \cite{calabrese2009entanglement}. At $t_1=t_2$, there exists a gapless point, which corresponds to a free Fermi sea that has two linearly dispersing Dirac fermions at $k = \frac{\pi}{2}$ and $\frac{3\pi}{2}$ (for a single site unit cell, the same points change to $k=0,2\pi$ for a two site unit cell due to Brillouin zone folding), rendering it a critical theory with central charge $c=1$. We now discuss the effects of including non-Hermiticity and the phases that emerge.

\subsection{$c=2$ chiral metal (CFS-A)}
\label{sec:cFSa}

The first non-trivial phase arises when we have both $t_1<t_2$ and $\gamma<t_2$. The Hermitian system at ($t_1\sim 0, \gamma=0$) exhibits two nearly flat bands, one of which is fully filled in equilibrium. Even a perturbatively small $\gamma$ immediately leads to a finite imaginary part of the energy eigenvalues [see \Fig{fig:dispersion1}(a)], which makes the equilibrium state unstable and takes the many-body state to a unique steady state which has a finite current.  It is in this sense that the non-Hermitian limit of the system is {\it not} perturbatively connected to the Hermitian physics. Not unlike the Hatano-Nelson model, where the system stabilizes a steady state with a Galilean boosted Fermi sea; here the Hermitian band insulator splits into two Fermi seas occupying the two bands (orange and green) respectively from $\pi<k <2\pi$ (orange band) and from $0<k<\pi$ (green) [see \Fig{fig:dispersion1}(a)].  

\begin{figure}[h!]
    \centering
    \includegraphics[width=1.0\columnwidth]{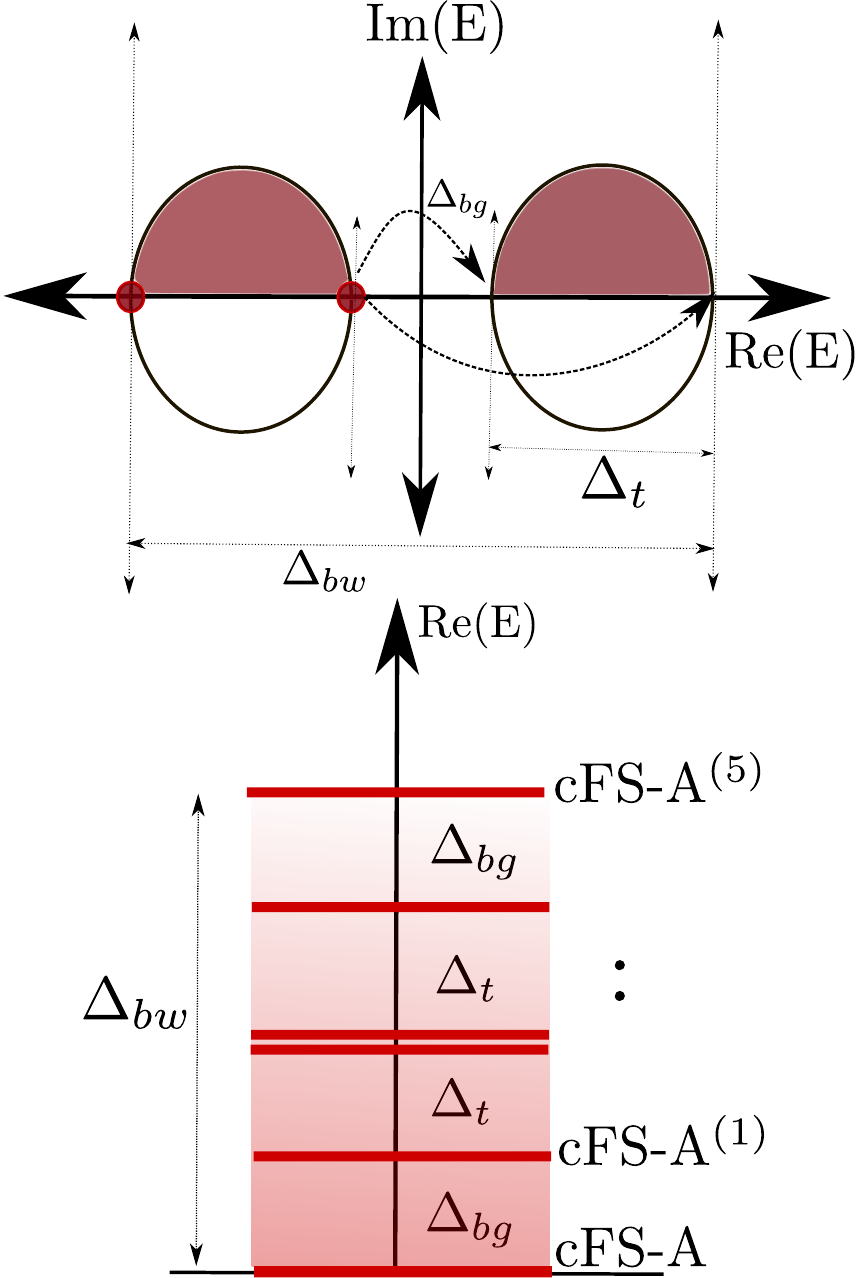}
    \caption{\textbf{Steady state and the nature of excitations in the chiral FS (A) phase.} A schematic showing the steady state as well as excitations in the cFS-A phase (see section.~\ref{sec:cFSa}). The spectral loops of width $\Delta_t$ (top) are separated by a line gap $\Delta_{bg}$ for a system of total bandwidth $\Delta_{bw}$. At half-filling six steady states exist (cFS-A$^{-(5)}$) which are separated in real energies due to particle hole excitations (shown using arrows). The six steady states are shown below.} 
    \label{fig:cFS2}
\end{figure}


{\it Nature of the steady state:} 
It is interesting to note that exactly at $\gamma=0$, the system has only the lower band filled. Inclusion of any finite $\gamma$ provides a time scale $\tau_{k,n}=\frac{1}{E^I_{k,n}}$ to each of the states after which the state grows (decays) given $E^I_{k,n}  > 0$ ($E^I_{k,n}< 0$). Interestingly, every band chooses both signs for $E^I_{k,n}$, therefore while the occupation of the lower band from $0<k<\pi$ is stable, at long times, the system moves to a steady state where the particles occupy the higher (real) energy band between $0<k<\pi$. 

An illustrative way of seeing this is to plot the single particle spectrum in the complex plane which shows two closed spectral loops [see \Fig{fig:cFS2} (top panel)]. These loops have a finite winding with a width $\Delta_{t}$ along the real axis. In the $\gamma=0$ limit this corresponds to the dispersing width of each of the bands. The loops are further separated in real energy by a ``band" gap of $\Delta_{bg}$ (also called line gap). The complete band width of the spectrum along real axis is $\Delta_{bw}$.

The existence of the spectral loops is symmetry protected, which guarantees existence of a state with complex energy $-E$ given a state with energy $E$. Given the existence of four eigenvalues that have a zero imaginary component, at half filling, the system has {\it six} many-body states which have the same total imaginary component, while different values of the real component. Therefore, both the spectral winding and the number of steady states in the system are intricately related. The non-equilibrium steady state (NESS) corresponds to the minimum total real energy and the maximum total imaginary component characterized by a split Fermi sea on both the lower and the upper band (see \Fig{fig:cFS2}) and four Fermi points.

{\it Excitations:} The NESS can excite into other five steady states (cFS$^{(1)-(5)}$) via particle-hole excitations, which leads to states that have the same total imaginary component, however having a gap in real energies determined by $\Delta_t$ and $\Delta_{bw}$ [see \Fig{fig:cFS2} (bottom)]. These excitations {\it do not decay} and can have finite amplitudes at long times. Each of these multiple steady states have indistinguishable current and real space correlations. However, these steady states themselves are {\it gapless} and are dominated by particle-hole excitations within the spectral loops which, in turn, are characterized by point gaps in the complex energy plane. Given the continuum of excitations within each loop, these excitations live longer, closer they are to the Fermi points.

{\it Current:} The occupied bands in the system lead to a finite current for the NESS, given that the single particle states are current carrying states and contribute additively to the many-body state. This is in contrast with any Hermitian system where current contributions from various single particle states cancel pairwise, leading to an equilibrium state with zero current. For instance, here both the $+$ band and the $-$ bands have a positive contribution to the current [see \Fig{fig:dispersion1}(b)]. It is easy to integrate $j_{k\alpha}$ to find the total current in the system, even when $\gamma$ is perturbatively small. This finite current

\beq
j_o = \frac{2}{\pi}t_1,
\eeq

is characteristic of the NESS. Non-zero $\gamma$ leads to a quadratic increase [see \Fig{fig:cfs-A} (a)]

\beq
J_{|\gamma| > \epsilon, \epsilon \rightarrow 0  } \sim \signum(\gamma) \frac{2 t_1}{\pi} (1+ \frac{3}{25}\gamma^2)
\label{c=2-current}
\eeq

In fact the same behavior is expected for all the six steady states in the system, given the current is determined collectively by the bulk of the single particle states.

Let us stress that the origin of this finite current in the NESS is purely topological in nature. A finite winding of $H(k)$ around any reference point in the complex spectral plane ensures the existence of a finite additive current from various single particle states, as seen in eqn. (\ref{Eq:TopCurrent}) \cite{Zhang_2020}. In the cFS-A phase, the current contributions stem from the winding of the two spectral loops. While the spectral topology of the Hamiltonian gives rise to finite current, the notion of steady state filling ensures a purely real total current through cancellation of all imaginary parts [see Fig.\ref{fig:dispersion1}(b)].

{\it Correlations:} Having discussed the current carrying nature of this state, it is natural to pose the question regarding how the fermions are mutually correlated. It is interesting to note that even while the fermions at the same sublattice are un-correlated, i.e., 

\begin{equation}
C^{A,A}(r) = C^{B,B}(r) = \frac{1}{2}\delta(r),
\end{equation}
the fermions on two different sublattices are power law correlated (to leading order dependence on $\gamma$)
\begin{equation}
C^{A,B}(r) \sim \frac{\gamma}{r} \sin( \frac{\Delta k_F r}{2}),
\label{corr-cfs-2}
\end{equation}
with characteristic oscillations governed by a finite density of fermions in each band ($\Delta k_F=\pi$). This $\gamma$ dependent behavior is in addition to a $\gamma$ independent power law with similar oscillations. This is in direct contrast with the Hermitian limit of a band insulator where each of these correlators fall exponentially with distance [see \Fig{fig:cfs-A}(b)]. It is also important to note that generally for a non-Hermitian system $\langle c^\dagger_i c_j \rangle \neq [\langle c^\dagger_j c_i \rangle]^\dagger $.  The above features, therefore, qualifies that this non-Hermitian phase is in fact a chiral-metal.

\begin{figure}
    \centering
    \includegraphics[width=1.0\columnwidth]{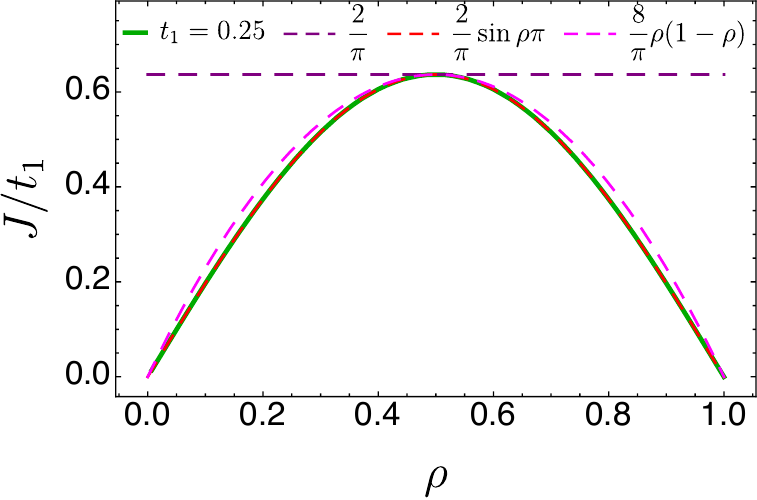}
    \caption{\textbf{Steady state current in cFS-A phase.} Steady state current, $J$, as a function of density, $\rho$, at $t_1=0.25$ in the cFS-A phase. Note the behavior characteristic of exclusion processes with a maximum at half-filling. Here we set $\gamma=0.1$. }
    \label{fig:currentcfs2dens}
\end{figure}

{\it Entanglement:}  
We next focus on the entanglement properties of this phase, where a part of the system under consideration in real space ($l$) is traced over, and the entanglement of the remaining part is calculated. Given the non-interacting system, we use the prescription given by Peschel and Calabrese \cite{calabrese2009entanglement, chang2020entanglement, guo2021entanglement} and extract the central charge of the system using (for periodic boundary conditions)

\begin{equation}
S = \frac{c}{3} \ln{[\frac{L}{\pi}\sin{(\dfrac{\pi l}{L}})]},
\end{equation}

where we find that $c=2$ [see \Fig{fig:cfs-A}(c)]. This is consistent with the fact that the system behaves like two disjoint Fermi surfaces (corresponding to two bands) with a Fermi volume of $\pi$, and therefore gapless low energy excitations on {\it four} Fermi points, where each potentially contributes $c=\frac{1}{2}$ to the central charge.

{\it Away from half-filling:} All of the discussion above have made use of the fact that we keep the non-Hermitian system at a finite density of fermions which is fixed to ${\it half}$. This implicitly assumes that even while single particle states themselves can decay, the system pumps in new particles to maintain a finite filling. However, a finite current is rooted in the spectral topology and the filling only changes the contribution of single-particle states to the total current. Away from half-filling the nature of the metallic state does not change, since both the bands still get partially but equally filled, the four Fermi points still remain, and correlations are still power law, albeit with a different wavelength of oscillation, which, in turn, corresponds to the changing Fermi volume. However the value of the steady state current is density dependent and follows a behavior characteristic of exclusion processes  $\sim  \rho(1-\rho)$ (see for e.g.~\cite{Derrida2007})  as seen in \Fig{fig:currentcfs2dens}. In this way it lends credence to the ideas of non-Hermitian systems modelling either open systems or boundaries of higher dimensional symmetry protected topological systems \cite{LeePRL2019}, where the boundary states are maintained at a finite filling even the while the system is connected to the leads.

\begin{figure}
\includegraphics[width=1.0\columnwidth]{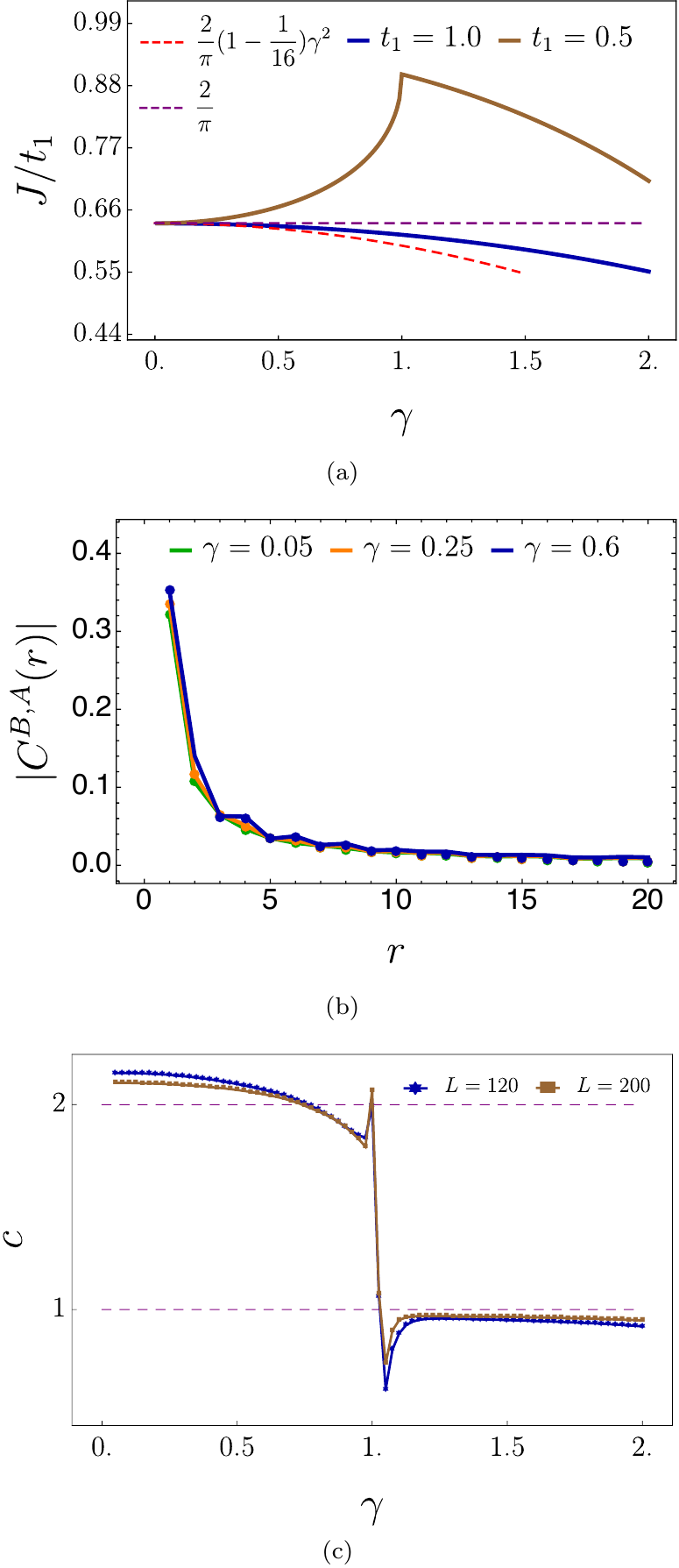}
   \caption{\textbf{Many body properties of $c=1$ cFS.} (a) Chiral current for two different values of $t_1$, as a function of the non-Hermitian co-efficient $\gamma$ [see Eq.\ref{c1Current}] (b) The correlator, $C_{B,A}(r)$, as a function of $r$ for $t_1=1.0$ for different values of $\gamma$. The numerical values from real space calculation are the points ($L=300$). The continuous lines are the expressions shown in eqn. (\ref{corrcfs1}) for appropriate parameters. (c) The central charge variation is shown as a function $\gamma$ ($t_1=0.5$) for two system sizes. The non-analytic behaviour in the current and the central charge arise at $t_1=0.5$ due to cFS-A to $c=1$ cFS transition.}
   \label{fig:c1cfs}
\end{figure}


\subsection{$c=1$ chiral metal (cFS)}
\label{sec:cFS1}

There exists another distinct chiral metallic phase in this system with two Fermi points. As we show below this phase is smoothly connected to the Hatano-Nelson steady state, which is realized in the celebrated single band non-reciprocal hopping Hamiltonian \cite{hatano1996localization}. The ground state of $t_1=t_2, \gamma=0$ point is a half-filled Fermi sea with two Fermi points. Any non-reciprocal hopping immediately shifts the Fermi sea to a finite-momentum current carrying steady state still with two Fermi points, and gapless excitations. While at $\gamma=0$ this phase is unstable to dimerization, a finite $\gamma$ opens up a window where even upon introduction of dimerization ($t_1\neq t_2$) the chiral metallic phase continues to remain stable. We next discuss the properties of this phase.

{\it Nature of steady state and excitations:} At $\gamma=0$ the complex spectrum is a single continuous line which spans the total bandwidth. The equilibrium state at half-filling is given by occupation of all single particle states with $Re[E]<0$. Introduction of a non-zero $\gamma$ adds an imaginary component to the single particle eigenvalues which leads to single loop in the complex spectrum (see Fig.\ref{fig:chiral-modes}). The existence of two eigenvalues with zero imaginary component leads to two steady states which are separated in real energy by their band width $\Delta_{bw}$. Each of these states however has a set of gapless excitations, which decay with time. 

{\it Current and correlations:} In this phase, both the steady states carry a finite current (see \Fig{fig:c1cfs}), where the current is given by

\beq
J \sim \signum{(\gamma)}\frac{2t_1}{\pi} [1-\frac{\gamma^2}{16}].
\label{c1Current}
\eeq

The current comes from the finite winding of a single loop in the complex spectral plane. As is expected of a gapless state the fermionic correlations decay in a power law fashion and the existence of a single Fermi surface with a Fermi volume of $2\pi$ leads to no characteristic oscillations in the inter-sublattice correlator such that even while $C^{A,A}(r) = C^{B,B}(r) = \frac{1}{2}\delta(r)$, one finds (the leading order dependence on $\gamma$)as
\begin{equation}
C^{A,B}(r) \sim \frac{\gamma}{r}.
\label{corrcfs1}
\end{equation}
which again is in addition to an $\gamma$ independent power-law part.

{\it Entanglement and density dependence:} Given two Fermi points, we expectedly find $c=1$ from the subsystem entanglement scaling (see \Fig{fig:c1cfs}), characterizing the phase as a $c=1$ chiral metal. Going away from half-filling again leads to a change in the current density, additionally resulting in oscillations in the correlators characteristic of the reduced Fermi volume (not shown).

\subsection{Entrapped band insulator}

\begin{figure}
\includegraphics[width=1.0\columnwidth]{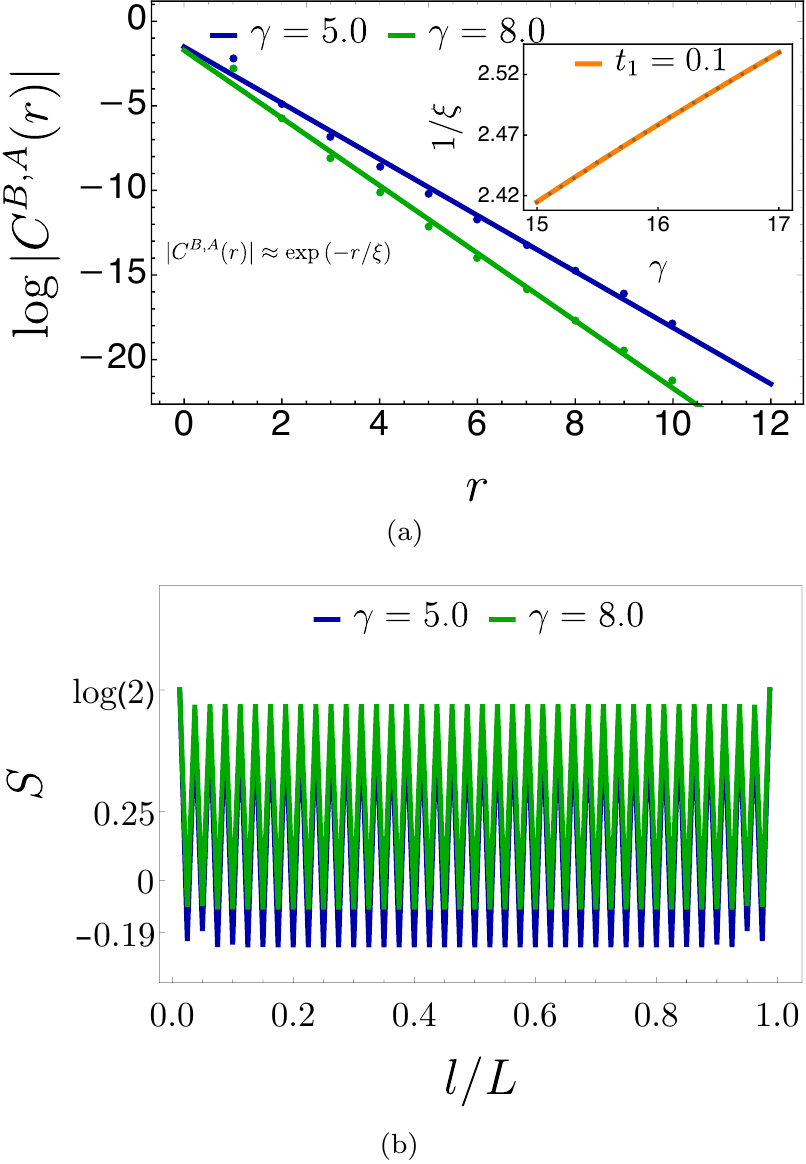}
\caption{\textbf{Many body properties of $c=0$ EI.} (a) The correlator $C_{B,A}(r)$ in logarithmic scale as a function of $r$ for different values of $\gamma$. The numerical values from real space calculation are the points ($L=300$). The continuous lines are the expressions shown in \eqn{eq:corrcI} for appropriate parameters. (b) Entanglement entropy for different parameter values of $\gamma$, as a function of $l/L$. Here, we have set $t_1=0.1$.} 
 \label{fig:cI}
\end{figure}


A novel non-equilibrium phase of matter which is realized here is that of an entrapped insulator (EI) which gets stabilized for $\gamma/t_2 > 2(t_1/t_2 +1)$  region of the phase diagram (see \Fig{phasediag}). One illustrative limit to start from is the $t_2=0$ ($t_1 \rightarrow \infty$) and $\gamma \neq 0$ where the chain breaks into a sequence of dimer states, albeit with strong non-reciprocal coupling within each dimer. At exactly half-filling, each unit cell can host a single electron stabilizing a non-Hermitian insulating state. Given every particle is dynamically trapped in a unit cell with a finite intra-unit cell current, we call this new state an EI. However, the inter-unit cell current remains zero. It is this phase which continues to remain stable for large values of $\gamma$ and a window of $t_1$ and $t_2$ regions.
 
{\it Nature of the steady state:} The complex spectrum again shows a two-lobe structure with a line-gap in this phase, however the line gap now exists in the imaginary eigenvalue direction (see \Fig{fig:chiral-modes}). At half-filling the upper lobe is completely filled creating a gap to excitations which are short-lived with a finite lifetime. 
 
{\it Current and correlations:}
Given the gapped character of the system this steady state has zero current. This is also seen through the fact that the spectral loop in this case has zero winding. The inter-sublattice fermion correlator decays exponentially with a form given by

\beq
C^{A,B}(r) \sim \exp(-r/\xi),
\label{eq:corrcI}
\eeq

where $\xi \approx 1/\gamma $ (see \Fig{fig:cI}).

{\it Entanglement:} The bipartite entanglement entropy as a function of subsystem size shows characteristic oscillations with periodicity of two sites signalling a dimer order with a value $\sim \log(2)$ when the partition is intra-unit cell, and $\sim 0$ when the partition is inter-unit cell (see \Fig{fig:cI}). One finds the inter-unit cell entropy to be negative for for larger values of $t_1$, as has been reported in other non-Hermitian systems \cite{chang2020entanglement}, however its interpretation in terms of information is not clear at present and merits further investigation.

\subsection{$c=2$ chiral metal (cFS-B)}
\label{sec:cFSb}

We now briefly discuss another chiral metallic phase which appears when $t_1 \gg t_2$. While many of the essential properties of this phase are similar to cFS-A, one distinctive feature is the orbital character of the steady state which we discuss next. 

At $\gamma=0$ the $t_1>t_2$ and $t_1<t_2$ phases are both insulating with a band gap, however the orbital character of the bands are distinct \cite{Asboth_springer_2016}. For instance for $t_1<t_2$ the half-filled equilibrium state has a distinct polarization compared to $t_1>t_2$ which is reflected in a winding number that changes discontinuously at $t_1=t_2$ \cite{Resta_book_2007, Resta_RMP_1994, Watanabe_PRX_2018}. Non-zero $\gamma$ does not change the orbital character of the single particle bands in the $\gamma=0$ limit. For instance, decomposing the band wavefunctions ($n=0,1$) into the bonding (antibonding) contributions ($c^\dagger_{nk} = \psi^n_{\alpha k} c^\dagger_{k\alpha} + \psi^n_{\beta k} c^\dagger_{k\beta} $ ) where
$c^\dagger_{k\alpha}  =  \frac{1}{\sqrt{2}} \Big(c^\dagger_{kA} + c^\dagger_{kB} \Big)$ and 
$c^\dagger_{k\beta}  =  \frac{1}{\sqrt{2}} \Big(c^\dagger_{kA} - c^\dagger_{kB} \Big)$, shows the characteristic twist of the bands when $t_1<t_2$ (when $\gamma \neq 0$) when $|\psi^n_{\alpha k}|^2$ is plotted for both the bands (see \Fig{fig:orbital-exchange}). Introduction of non-Hermitian instability due to $\gamma$ leads to half-filling
of both these bands (therefore maximizing the total imaginary component) for the steady state.

Moreover, the current in the system is dominated by the lower hopping scale which is $t_2=1$ in the regime when $t_1>t_2$.  Therefore the current in cFS-B starts at $2/\pi$ when $\gamma \rightarrow 0$ and is independent of the value of $t_1$ (See Fig. \ref{fig:Currentt1}).

\begin{figure}
\centering
\includegraphics[width=1.0\columnwidth]{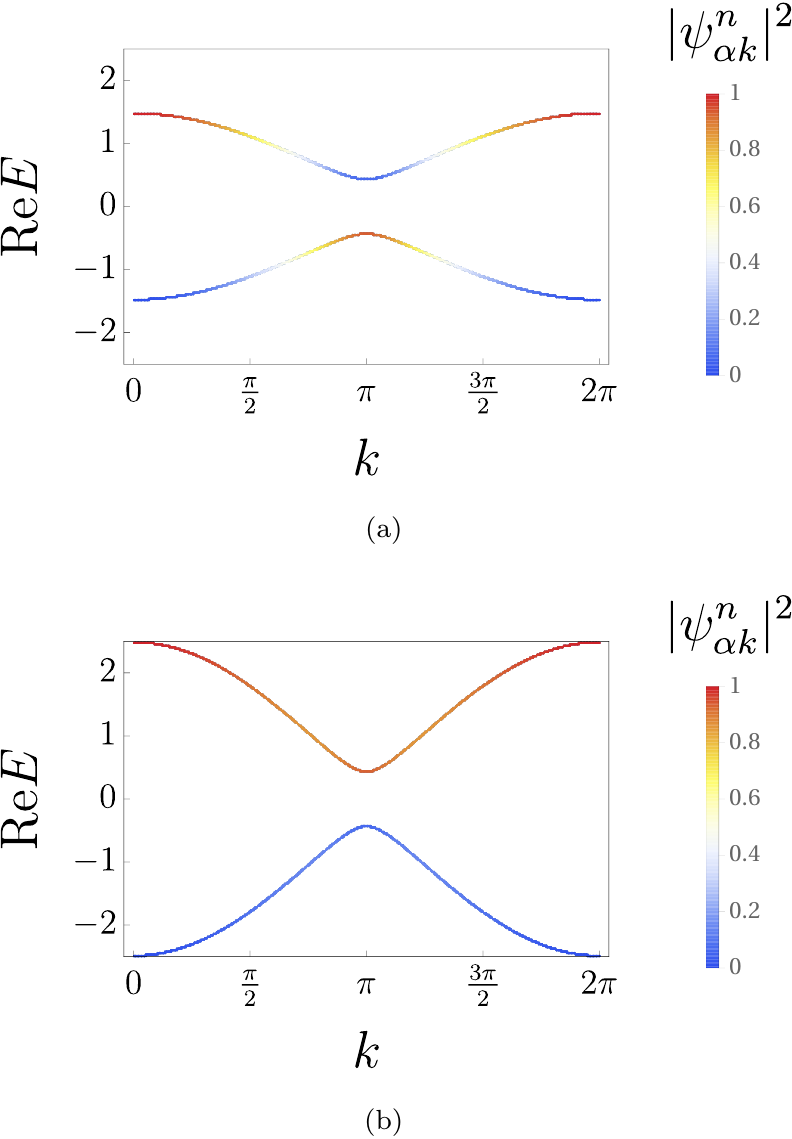}
\caption{\textbf{Distinction between two chiral phases cFS-A and cFS-B through orbital band character.} The real part of the dispersion is shown with $|\psi^n_{\alpha k}|^2$ contribution in color [see \sect{sec:cFSb}] with (a) $t_1=0.5$ and $\gamma=0.5$, and (b) $t_1=1.5$ and $\gamma=0.5$. Note the difference in the orbital character around $k=\pi$ for the two cases.} 
 \label{fig:orbital-exchange}
\end{figure}


\begin{figure}
\includegraphics[width=1.0\columnwidth]{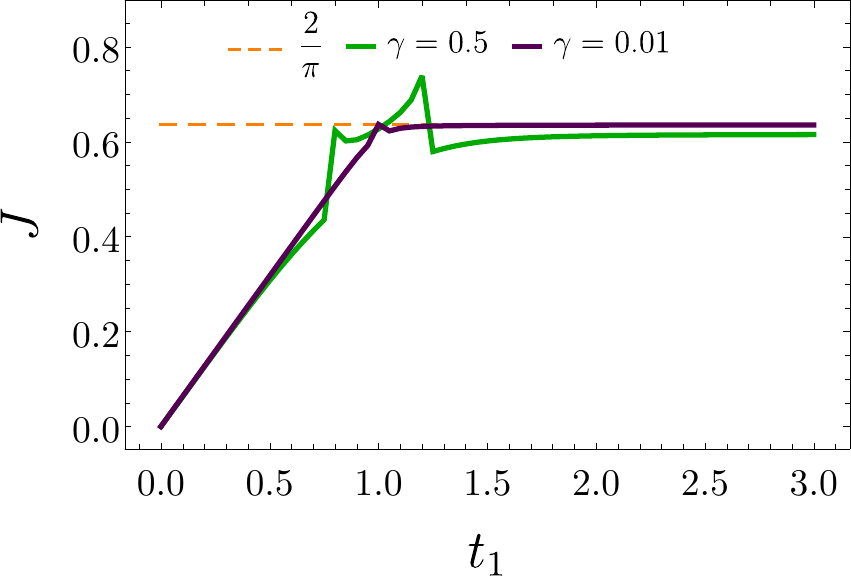}
 \caption{\textbf{Scaling of chiral current.} Chiral current as a function of $t_1$ for different values of non-Hermitian coefficient $\gamma$. The current in the system saturates to $2/\pi$ as a function of $t_1$ for ($t_1 \geq 1.0$). The current is  dominated  by  the  lowest of the two hopping  scales $t_1,t_2$.}
 \label{fig:Currentt1}
\end{figure}


\subsection{Low energy theories and nature of phase transitions}

\begin{figure}
    \centering
    \includegraphics[width=1.0\columnwidth]{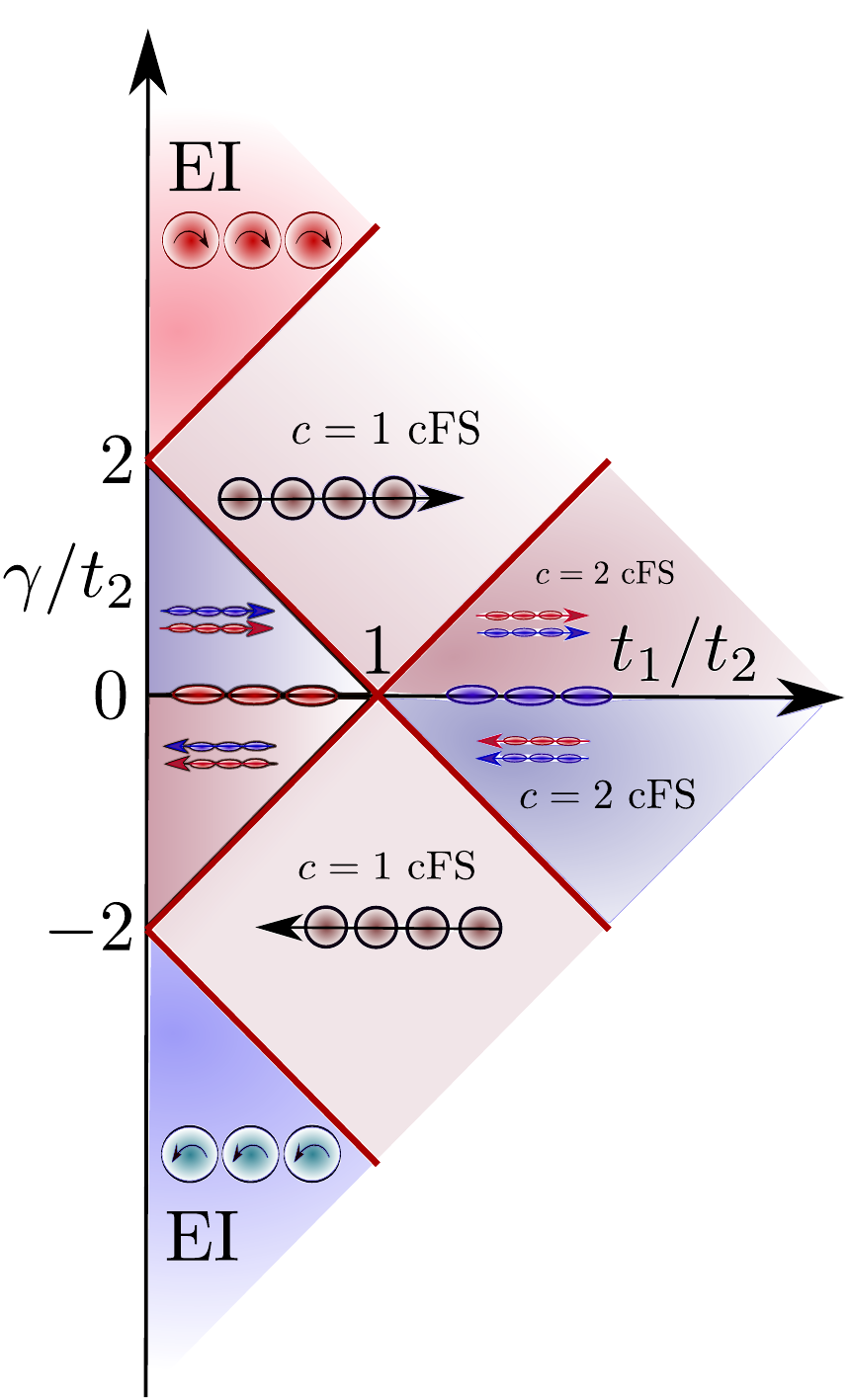}
    \caption{\textbf{Schematic of extended phase diagram.} A schematic of the extended phase diagram of the system showing different non-equilibrium phases[see \eqn{eqn:Ham}] including the $c=2$ cFS, $c=1$ cFS and the EI phase.}
    \label{PDphases}
\end{figure}


Having the discussed the character of the steady states and their low energy excitations, we now discuss the nature of these phase transitions between various chiral metals and entrapped insulating states (\Fig{PDphases}). We note that a conventional energy minimization to find the ground state or the excitations is not sufficient to describe the steady states. Therefore, one needs to go beyond conventional paradigm to describe such phase transitions, as we will see next. 

\subsubsection{Hatano-Nelson Model and $c=1$ cFS}

The Hatano-Nelson Hamiltonian is given by~\cite{hatano1996localization}

\beq
H_{HN} = \sum_i -t ( c^\dagger_{i} c_{i+1} + h.c.) + \gamma ( c^\dagger_{i} c_{i+1} - h.c.),
\eeq

which leads to a single particle dispersion given by $\epsilon(k) =  -2t\cos(k) +  2\ci\gamma \sin(k)$. While for the equilibrium Hermitian system ($\gamma=0$), the ground state is given by $0<k<\frac{\pi}{2}, \frac{3\pi}{2}<k<2\pi$, which has an entanglement scaling $c=1$ with low energy states that disperse linearly $
H(k) \sim v_F |k| \Big(  c^{\dagger R}_kc^R_k  + c^{\dagger L}_kc^L_k \Big),$
where $R,L$ specify the right and left moving modes.

In presence of any $\gamma$ the single particle eigenvalues get a positive imaginary coefficient for $\pi<k<2\pi$ leading to a steady state where $\pi<k<2\pi$ are occupied.


\begin{figure}
    \centering
    \includegraphics[width=1.0\columnwidth]{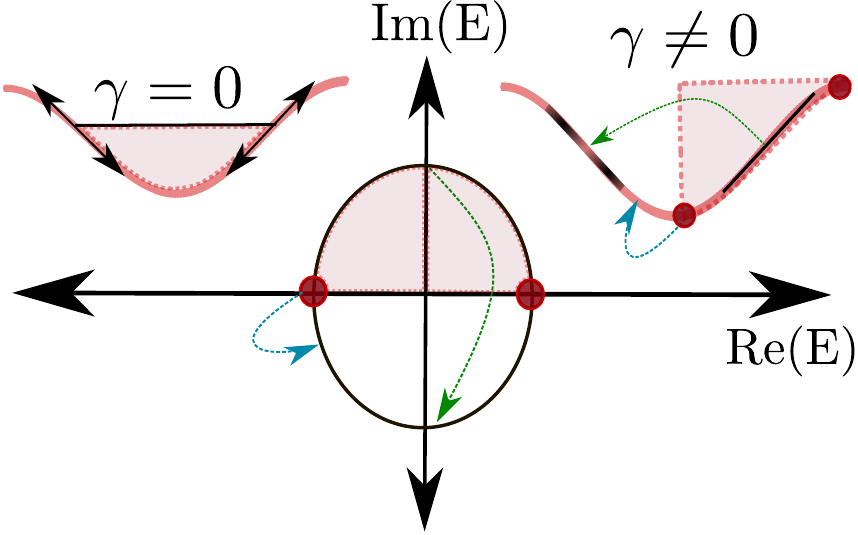}
    \caption{\textbf{Hatano-Nelson model at half-filling.} In the absence of non-Hermiticity ($\gamma=0$) the equilibrium system is given by two Fermi points with linearized excitations. In presence of finite $\gamma$ the shifted Fermi sea has quadratic excitations at band minima and band edges (blue), and short-lived linear excitations (green).}
    \label{fig:Hnelson}
\end{figure}

The low energy excitations is therefore {\it quadratically} dispersing fermions, albeit with a lifetime that is proportional to $\frac{1}{k}$,

\beq
H_{HN}(k) \sim \Big[ (-2 + k^2)t -\ci 2 \gamma k \Big] c^{\dagger R}_kc^R_k  +  \Big[ (2-k^2) -\ci 2 \gamma k \Big] c^{\dagger L}_kc^L_k.
\eeq

Along with excitations, another branch of short-lived excitations exist where the linearly dispersing fermions can scatter from the right (left) moving branch to the other left (right) branch with an energy $|(k-k_o)|$ ($k_o=\pi$) which are {\it linear} in momentum but with a finite lifetime (see \Fig{fig:Hnelson}). These excitations carry current in the opposite direction but eventually decay, thereby not changing the total steady state current of the many-body state. Interestingly the sub-system scaling of the entanglement entropy continues to show $c=1$. It is this feature which is characteristic of the chiral Fermi sea and is shared with the $c=1$ chiral Fermi surface which occurs in the complete phase diagram of \Fig{phasediag} for the nH SSH model.

In a reduced Brillouin zone for a two site unit cell, as is appropriate for our system, the above picture continues to hold, however, with both the effective mass of fermions and lifetimes being $\gamma$ and $t_1$ dependent.

\beq
H(k) \sim \Big[ (-\epsilon + \frac{k^2}{m} ) +\ci \frac{ k}{\tau} \Big] c^{\dagger R}_kc^R_k  +  \Big[ (\epsilon-\frac{k^2}{m}) +\ci \frac{k}{\tau} \Big] c^{\dagger L}_kc^L_k,
\eeq
where $\epsilon=2+t_1-\frac{\gamma^2}{16}, m^{-1}=(1/4)+\frac{t_1}{8}-\frac{\gamma^2}{128}$ and $\tau=\frac{8}{(2-t_1)\gamma}$ about $\gamma=0,t_1=1$. Interestingly, the short-lived linear excitations now are also at small momentum given the zone-folding with a velocity scale given by $\sim t_1$ and a $\tau \propto \frac{1}{\gamma}$.

\subsubsection{$c=2$ cFS to $c=1$ cFS}

The $c=2$ cFS is essentially two copies of the above Hatano-Nelson phase, however, made out of bonding and anti-bonding orbitals of the dimerized system, which are at two different values of real energies.
The natural point to investigate the $c=2$ FS is to start from $t_1=0, \gamma=0$ point, where the Hamiltonian is exactly diagonalized by the following transformation

\bea
c_{i\alpha} = \frac{1}{\sqrt{2}} ( c_{i+1,A} + c_{iB} ), \quad
c_{i\beta} = \frac{1}{\sqrt{2}} ( c_{i+1,A} - c_{iB} ).
\label{orbital}
\eea

A perturbative $\gamma$ when projected to just the $\alpha$ and $\beta$ bands to leading order gives the following non-Hermitian Hamiltonian

\bea
H_\alpha = \sum_i - ( c^\dagger_{\alpha i} c_{\alpha i} ) + \frac{\gamma}{4} ( c^\dagger_{\alpha i} c_{\alpha, i+1} - h.c.), \\
H_\beta = \sum_i + ( c^\dagger_{\beta i} c_{\beta i} ) + \frac{\gamma}{4} ( c^\dagger_{\beta i} c_{\beta, i+1} - h.c.), 
\label{cFS2eff}
\eea
where we have ignored the terms which mix $\alpha,\beta$ sectors. The above effective Hamiltonian essentially leads to {\it two} Galilean-boosted Fermi seas, which however, are separated in real energies (see \Fig{fig:cFS2}). A change of sign of $\gamma$ leads to a different $c=2$ cFS, where the currents are in the opposite direction. A similar interpretation is valid for large $t_2$, where $t_1 \rightarrow 0$ and another set of dimers can be formed with $zero$ polarization, $c_{i\alpha} = \frac{1}{\sqrt{2}} ( c_{i,A} + c_{iB} )$ and $c_{i\beta} = \frac{1}{\sqrt{2}} ( c_{i,A} - c_{iB} )$, with the effective non-Hermitian model given by \eqn{cFS2eff} with a different orbital content. Again for $\gamma<0$ the current direction changes. Therefore one realizes four different kinds of $c=2$ cFS (see \Fig{PDphases}) in the extended regime of $\gamma$ and $t_1$. Just like the discussion of the single Hatano-Nelson model in the last subsection, here for each $c=2$ cFS, one obtains two copies of quadratic and linear excitation branches.

The transition from such a $c=2$ cFS to a $c=1$ cFS is essentially a Lifshitz transition characterized by exceptional point physics at $k=\pi$. The eigenvalues of the single particle Hamiltonian at $k=\pi$ is given by [see \eqn{specexpan}] $E_k = \pm \sqrt{(t_1-t_2)^2 - \frac{\gamma^2}{4}}$, which shows that while the real spectrum is gapped as soon as $t_1 \neq t_2$ for $\gamma=0$, there exists an extended regime for $\gamma > 2|(t_1-t_2)|$ where the real eigenvalues are gapless, while the imaginary eigenvalues show a gap. Therefore, starting at any $t_1, t_2$ ($\gamma \rightarrow 0$) where a cFS-2 phase exists $\Delta_{bg} \sim |t_2-t_1| $ (see \Fig{fig:cFS2}) with quadractically dispersing fermions. Upon tuning the non-Hermiticity parameter, $\gamma$, the branches merge at a critical value of $\lambda_c = 2|(t_1-t_c)|$ close to which $\Delta_{bg} \sim \sqrt{\gamma_c-\gamma}$ -- showing the characteristic non-analyticity of a phase transition governed by exceptional point physics. Interestingly both the real and imaginary eigenvalues of energy scale as $\sim \sqrt{k}$ reflecting the non-analyticity of this low energy theory at the exceptional points \cite{aron2020landau}.

\subsubsection{Entrapped insulator to $c=2$ cFS}

To understand the entrapped insulator, it is illustrative to discuss the large $\gamma$ limit, where a different band insulator can be obtained with eigenfunctions
$\frac{1}{\sqrt{2}} \Big( \pm \ci c^\dagger_{iA} + c^\dagger_{iB}\Big)$ -- this has a finite {\it intra-unit cell} current but no inter-cellular current. The transition from the $c=2$ cFS ($t_1<t_2, \gamma \rightarrow 0$) to an insulating steady state is again by an exceptional point, where the imaginary gap scales as $\sqrt{\gamma}$ near the phase transition.

\begin{figure}
    \centering
    \includegraphics[width=1.0\columnwidth]{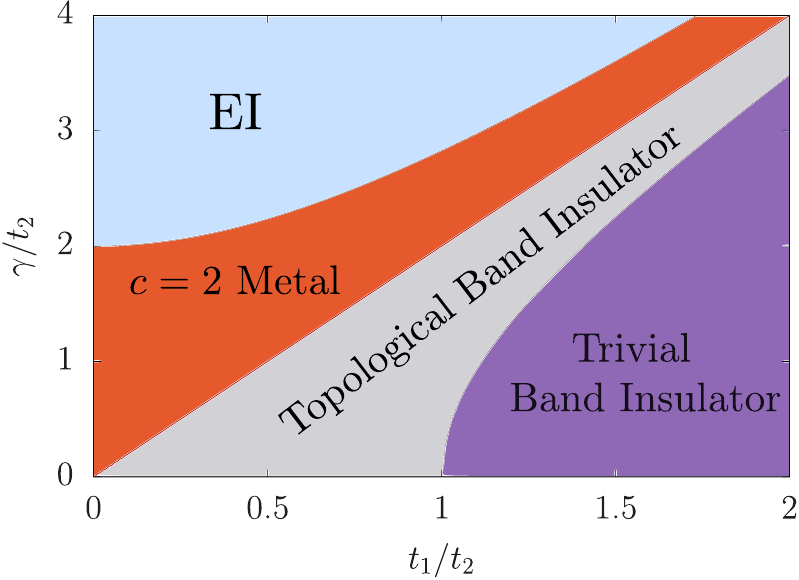}
    \caption{\textbf{Phase diagram under open boundary conditions.} Four phases arise in the non-Hermitian SSH model [see \eqn{eqn:transformation}] under open boundary conditions. Of these (i) entrapped insulator (EI) and (ii) $c=2$ metal, are non-equilibrium phases in their steady states, while (iii) topological band insulator and (iv) trivial band insulator are in equilibrium. }
    \label{phasediagopen}
\end{figure}


\begin{figure}
\includegraphics[width=0.85\columnwidth]{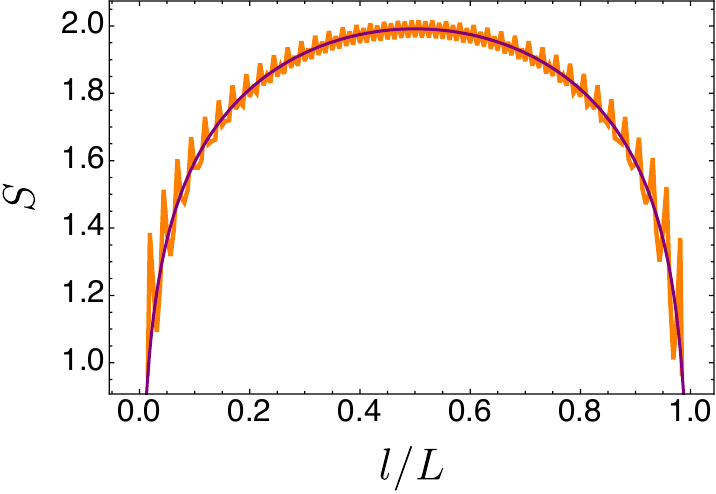}
\caption{\textbf{Entanglement entropy for $c=2$ metal.} The entanglement entropy for $c=2$ metal under open boundary conditions as a function of the bond length $l/L$ for $L = 160$. The numerical data can be fit  to $\frac{c}{6}\ln{[\frac{L}{\pi}\sin{(\dfrac{\pi l}{L}})]}$, where, $c/6 = 0.337$ (solid purple line). Here we have set $t_1=0.1$, $\gamma=0.75$.}
\label{cfs-a to topological band}
\end{figure}


\begin{figure}
\includegraphics[width=0.85\columnwidth]{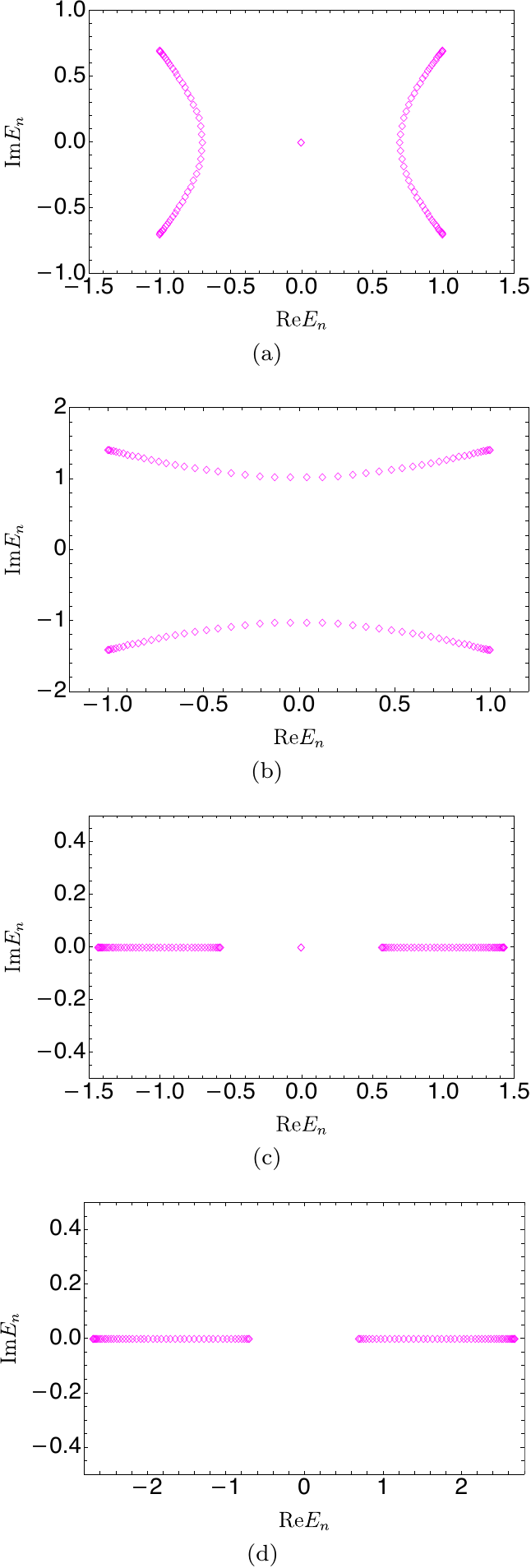}
\caption{\textbf{Single particle complex spectra under open boundary conditions.} (a) $c=2$ metal ( $t_1=0.5$, $\gamma=1.72$)  (b) Entrapped insulator( $t_1=0.5$, $\gamma=3.0$)  (c) Topological band insulator ( $t_1=0.5$, $\gamma=0.5$)  (d) Trivial insulator phase ( $t_1=2.0$, $\gamma=0.5$) }
\label{complex-obc}
\end{figure}


\section{Tuning boundary conditions}
\label{sec:boundary}

Having discussed the properties and phases of our system under periodic boundary conditions, we next study the system under open boundary conditions. For the Hermitian case, we know that an open system contains boundary modes which are robust to disorder. It is natural question to pose if such modes can be obtained in a non-Hermitian system and how they effect the steady states. 

More generally, we multiply a scaling factor, $\lambda$, to the boundary bond in order to smoothly tune between an open and a periodic problem (see \cite{koch2020bulk}), and discuss its effect on the steady state physics. When $\lambda=0$ (open boundary conditions), as has been reported in earlier studies focusing on single particle spectrum (see, for e.g., Reference~\cite{Herviou_PRA_2019}), one finds four distinct regions as shown in \Fig{phasediagopen}. Briefly, for $\gamma>2t_1$ all eigenvalues are real, while for $\gamma < 2t_1$ eigenvalues can have both real and imaginary components. Moreover, within the region $ 2\sqrt{t^2_1-1} < \gamma < 2\sqrt{t^2_1+1}$, one finds zero energy eigenvalues which smoothly connect to topological boundary modes in the Hermitian limit ($\gamma=0, t_1<t_2$). Given that our steady states are constructed by populating modes with the largest imaginary component, such long-lived non-equilibrium states are only defined for $\gamma<2t_1$, while for $\gamma>2t_1$ one obtains a conventional equilibrium ground state which is obtained by minimizing the total real energy. This apparent mismatch between the thermodynamic phase diagram for the periodic and the open problem is, however, an artifact of extreme sensitivity of the non-equillibrium phases to the $\lambda=0$ limit, and, as we show below, any infinitesmal $\lambda$ leads to stable steady states albeit with subtle effects arising due to topological boundary modes.

Four distinct phases can be identified in the open system as shown in \Fig{phasediagopen} and \Fig{complex-obc}. Out of these, two: (i) trivial band insulator and (ii) topological band insulator are the equilibrium phases which are smoothly connected to the Hermitian limit ($\gamma=0$). While in the thermodynamic limit both these regions have a bulk band gap, the latter has quasi-degenerate boundary modes at zero real energies. The other two phases (iii) $c=2$ metal and (iv) entrapped insulator, are non-equilibrium steady states which arise by populating the eigenmodes with largest imaginary components. We find that the metallic phase has gapless excitations which can be long-lived, and has a zero total current in the system (as is expected for an open system). Investigating the bipartite entanglement leads to a central charge $c=2$ (see \Fig{cfs-a to topological band}). Interestingly, this metallic phase contains two zero energy modes which have both real and imaginary components to be zero. This leads to quasi-degenerate steady states in the system which can be excited via particle-hole excitations. Having discussed the many-body phases as realized in the completely open system, we now investigate their stability to weak hybridization ($\lambda$) between the boundary sites.

We find that, in the thermodynamic limit, any perturbatively small $\lambda$ {\it immediately} leads to the non-equilibrium phases as realized in the periodic system as shown in 
\Fig{phasediag}. However, as $\lambda$ is tuned, two phenomena generically occur (a) boundary modes in regions of the metal and band insulator give way to  Fermi seas albeit with boundary modes either by changing the gap between the steady states or by leaky boundary modes, and (b) the steady state current gets immediately saturated to the value in the periodic problem as soon as $\lambda \neq 0$. We therefore interpolate from the open (see \Fig{phasediagopen}) to periodic system as a function of $\lambda$ (see \Fig{phasediag}) to see how the steady state evolves. In the Hermitian limit $(\gamma=0)$ when $t_1<t_2$ the boundary modes for the open system hybridize and gap out (with the gap $\propto \lambda$), eventually merging with the bulk bands at $\lambda=1$.

\begin{figure}
\includegraphics[width=1.0\columnwidth]{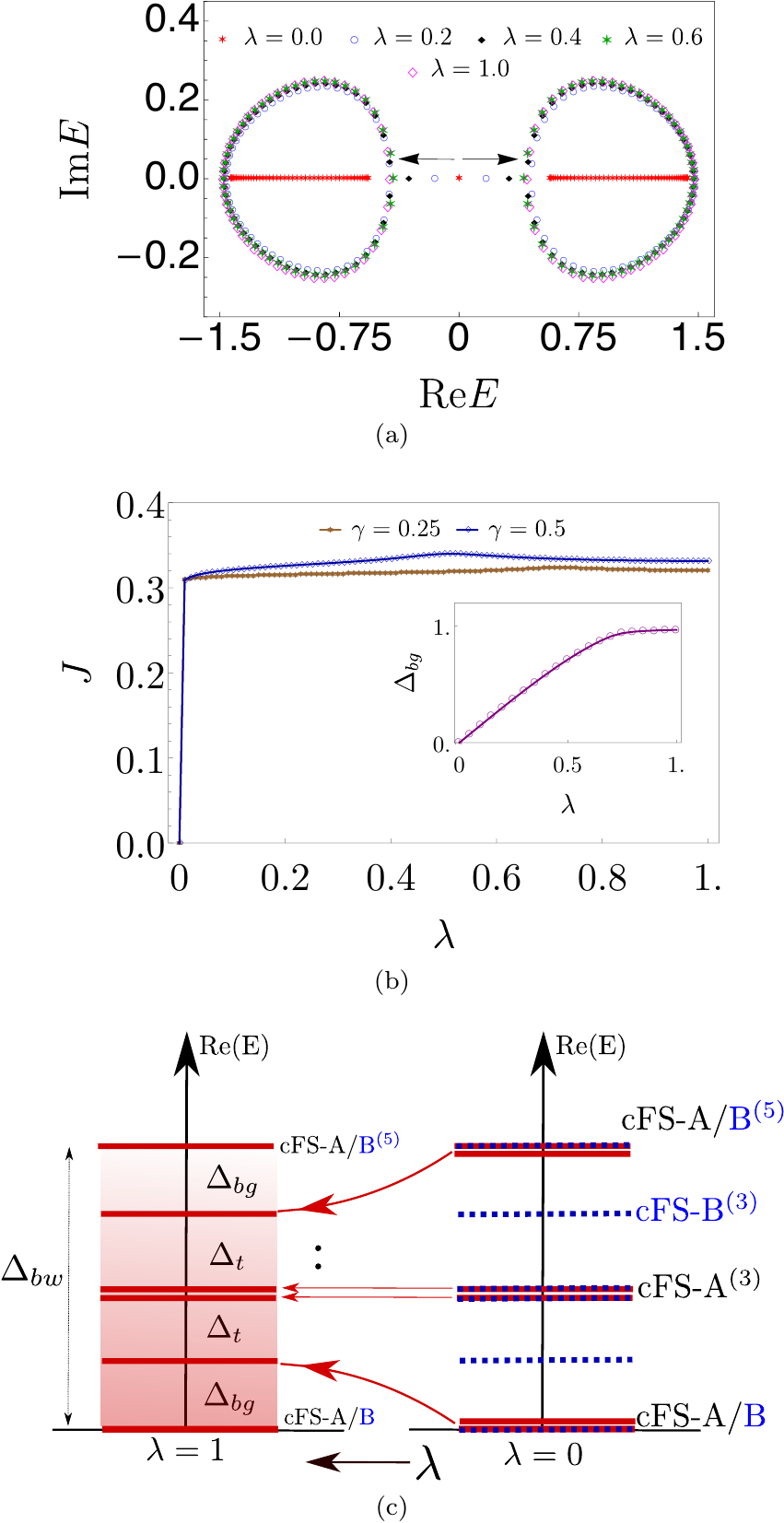}
\caption{\textbf{Instability of the topological band insulator region to cFS-A.} (a) The spectrum in the complex plane for different values of $\lambda$ for $t_1=0.5, \gamma=0.5$. (b) The current, $J$, as a function of $\lambda$. The inset shows the behaviour of $\Delta_{bg}$ by tuning $\lambda$. (c) A schematic of the energy spectrum. Here $t_1=0.5$.}
\label{TBIc2FS}
\end{figure}


\begin{figure}
\includegraphics[width=0.90\columnwidth]{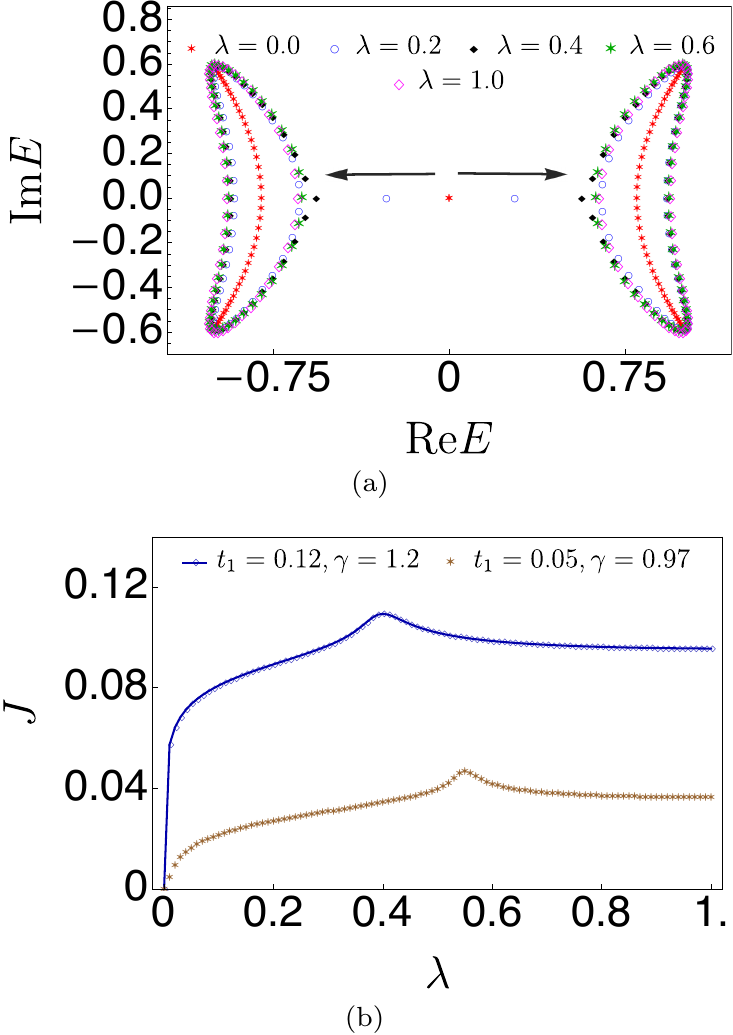}
\caption{\textbf{Instability of the $c=2$ metal region to $c=2$ cFS-A.} (a) The complex spectrum for different values of $\lambda$ for $t_1=0.12, \gamma=1.2$. (b) The current, $J$, variation with $\lambda$ for two different representative values of $t_1$ and $\gamma$. }
\label{c2tocfsA}
\end{figure}

\paragraph{Instability to cFS-2 A:}{We start from the open system in either the topological (or trivial) band insulator and introduce a $\lambda$ to interpolate to the $c=2$ cFS. While the spectrum is purely real for $\lambda=0$, any finite value of $\lambda$ introduces an imaginary part to the eigenvalues and moves the boundary modes along the real line towards the two lobes [see the spectrum in \Fig{TBIc2FS} (a)]. This immediately leads to a finite steady state current as shown in \Fig{TBIc2FS} (b) for a few representative values. Therefore, we find that the open boundary phase diagram is {\it not stable}, and immediately leads to a chiral metal phase. However, there is a distinction as to how the $c=2$ cFS is approached either from topological or trivial band insulator regimes. Note that the $c=2$ cFS has six steady states with gaps given by $\Delta_{bg}$ and $\Delta_{t}$ (see \Fig{fig:cFS2}). When $\lambda$ is tuned such that the $c=2$ CFS  approaches a topological band insulator, two bulk modes move towards \textit{zero} in the real spectrum thereby tuning both $\Delta_t \rightarrow \Delta_{bw}/2$ and $\Delta_{bg} \rightarrow 0$ [see \Fig{TBIc2FS} (c)], thus offering a tunability between many-body two quasi-degenerate steady states by occupying or emptying the boundary mode locally. It is in this sense that these steady states realized in $c=2$ cFS-A are in fact {\it tunable} and are inherently distinct from the $c=2$ cFS-B. Thus the role of single particle boundary modes here is to provide an avenue for interesting quasi-degenerate steady states which are boundary tunable. Similar physics is at play when $\lambda$ interpolates between $c=2$ non-chiral metal realized in the open system to the $c=2$ chiral FS. Again any finite value of $\lambda$ immediately leads to a chiral current, and existence of boundary modes, which, in turn, lead to quasi degenerate steady states (see \Fig{c2tocfsA}).
}


\begin{figure}
\centering
\includegraphics[width=0.90\columnwidth]{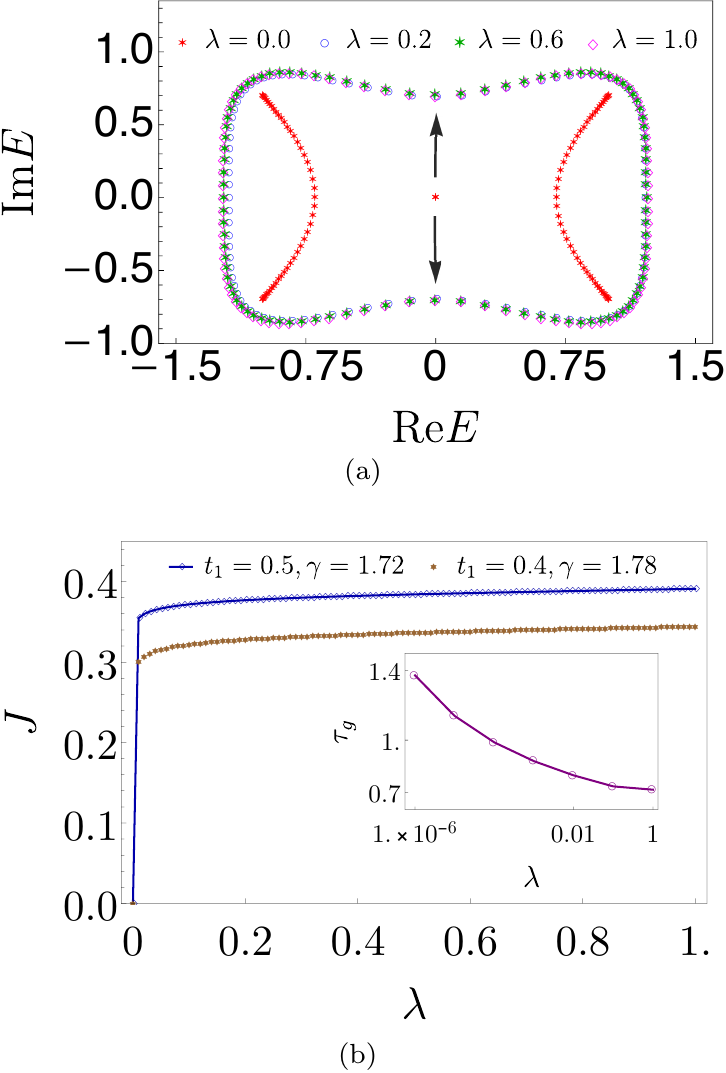}
\caption{\textbf{Instability of  the $c=2$ metal region to $c=1$ cFS.} (a) Tuning of the complex spectrum with varying $\lambda$. (b) Variation of the current, $J$, with $\lambda$ at two representative choices of $t_1$ and $\gamma$. (inset: variation of the lifetime $\tau_g$ of the boundary excitations as a function of $\lambda$ for $t_1=0.5, \gamma=1.72$).}
\label{c2metal}
\end{figure}


\paragraph{Instability to $c=1$ cFS:}{We next discuss the instability to another phase from the topological band insulator, where the system goes to the chiral $c=1$ FS. At $\lambda=0$, given that we are starting at the topological band insulator, we have two zero energy eigenmodes. However, unlike the instability to $c=2$ phase, here the boundary modes move along the imaginary axis leading to a  finite lifetime for the boundary mode (see \Fig{c2metal}). The system immediately obtains a finite current and two degenerate steady states. Each of them, however, has an excitation over the boundary which has a finite lifetime. The way the value of lifetime changes with $\lambda$ is shown in the inset of \Fig{c2metal} (b). Unlike the previous case, the lifetime decreases exponentially with increasing $\lambda$ leading therefore to the {\it leaky} boundary modes.}

The entrapped insulator phase in the open system gets smoothly connected to the periodic system and is therefore stable to tunability of boundary conditions as is expected for a steady state with no finite current. With this we have discussed the stability of the all the phases we realized in the open system and the role of boundary modes in such steady state physics.

\section{Summary and discussion}
\label{sec:summary}

In this work we revisited the nH-SSH model and the phases it realizes under finite filling. We find that the single-particle phase diagram gets reinterpreted in terms of distinct non-equilibrium phases of matter, which get realized at long times. We further analysed each of the phases in terms of its correlations, gap to excitations and entanglement features. Developing the low energy theories of various phases and intervening transitions, we build a qualitative understanding of the phases and its excitations in terms of the spectral topology. We further tuned the boundary conditions to take the system from an open to a periodic problem, and find that the nature of the equilibrium phases in the open system are perturbatively unstable to a weak coupling and leads to finite currents in the system.

An important question, that we have side-stepped here, is how such non-Hermitian Hamiltonians get realized in an experimental setting and where this is a valid effective description.
While in open systems, where bath degrees of freedom are systematically integrated out and the effective density matrix of a system is evolved -- one naturally uses the Lindbaldian approach which can in general be non-Hermitian \cite{AltlandPRX2021,bergholtz2021exceptional}. Here we presume the existence of the nH-SSH system and that one can keep it at a finite density. This is a particularly useful construct where one may consider such non-Hermitian models and their steady states to be anomalous theories which get realized on the boundaries of higher-dimensional topological phases \cite{LeePRL2019}.

Our work opens up several avenues to explore. One natural scope of expanding our findings is to consider higher dimensions, where one expects such chiral phases and entrapped insulating phases to be generic. An interesting direction will be to investigate the role of disorder and impurities in such systems and the meaning of resulting topological protection in such many-body states obtained in the non-Hermitian phases. How an equilibrium state reaches such steady states in time and the corresponding dynamics of the density matrix due to introduction of non-Hermiticity is another interesting future scope. Similarly the role of many-body physics and interactions is something that has been studied only recently\cite{alsallom2021fate,Lee_PRB_2020,Liu2020,Mu20,Shen21,zhang2021observation,Yoshida21} and is worth pursuing in future. In conclusion, our different many-body perspective on the non-Hermitian phases can lead to interesting insights where the interplay of topology, entanglement and correlations conspire to produce novel phases of quantum matter.

\acknowledgements
AB is supported by the Prime Minister's Research Fellowship (PMRF). SSH acknowledges MPI-PKS, Dresden for support during the project. AA acknowledges enlightening discussions with Animesh Panda, Sumilan Banerjee, Vatsal Dwivedi, Aabhaas Mallik, Abhisodh Prakash and a related collaboration with Amit Chatterjee and Arghya Das. AA acknowledges partial financial support through Max Planck partner group on strongly correlated systems at ICTS. AA acknowledge the visitor's program at MPI-PKS, Dresden for hospitality.  AN is supported by the start-up grant (SG/MHRD-19-0001) of the Indian Institute of Science and by DST-SERB (project number SRG/2020/000153).


\bibliography{references,ref1DCorr}

\end{document}